\documentclass[longauth]{aa}

\usepackage{amsmath}
\usepackage{amssymb}
\usepackage{graphicx}
\usepackage{graphics}
\usepackage{epsfig}
\usepackage{longtable}
\usepackage{blindtext}
\usepackage{float}
\usepackage{siunitx}
\usepackage{bm}

\usepackage{longtable}
\usepackage{threeparttable}

\usepackage{enumitem}
\usepackage[usenames, dvipsnames]{color}

\usepackage{txfonts}
\usepackage[pdfstartview=XYZ,
bookmarks=true,
colorlinks=true,
linkcolor=blue,
urlcolor=blue,
citecolor=blue,
pdftex,
bookmarks=true,
linktocpage=true,
]{hyperref}

\newcommand\T{\rule{0pt}{2.6ex}}       
\newcommand\B{\rule[-1.2ex]{0pt}{0pt}} 

\newcommand{\Lya}{Ly$\alpha$}
\newcommand{\Ha}{H$\alpha$}
\newcommand{\Hb}{H$\beta$}
\newcommand{\farc}{\hbox{$.\!\!^{\prime\prime}$}}

\newcommand{\kms}{\,km\,s$^{-1}$}
\newcommand{\Msunyr}{\hbox{M$_{\rm{\odot}}\,\rm{yr^{-1}}$}}

\newcommand{\ergscmA}{erg\,s$^{-1}$\,cm$^{-2}$\,\AA$^{-1}$}
\newcommand{\ergscm}{erg\,s$^{-1}$\,cm$^{-2}$}
\newcommand{\lognhicm}{$\log(N_{\rm H{\tiny I}}$/cm$^{-2})$}
\newcommand{\HI}{H{\tiny I}}
\newcommand{\NHI}{$N_{\rm H{\tiny I}}$}

\usepackage{natbib,twoopt}
\bibpunct{(}{)}{;}{a}{}{,}            
\makeatletter
  \newcommandtwoopt{\citeads}[3][][]{\href{http://adsabs.harvard.edu/abs/#3}%
    {\def\hyper@linkstart##1##2{}%
     \let\hyper@linkend\@empty\citealp[#1][#2]{#3}}}
  \newcommandtwoopt{\citepads}[3][][]{\href{http://adsabs.harvard.edu/abs/#3}%
    {\def\hyper@linkstart##1##2{}%
     \let\hyper@linkend\@empty\citep[#1][#2]{#3}}}
  \newcommandtwoopt{\citetads}[3][][]{\href{http://adsabs.harvard.edu/abs/#3}%
    {\def\hyper@linkstart##1##2{}%
     \let\hyper@linkend\@empty\citet[#1][#2]{#3}}}
  \newcommandtwoopt{\citeyearads}[3][][]%
    {\href{http://adsabs.harvard.edu/abs/#3}
    {\def\hyper@linkstart##1##2{}%
     \let\hyper@linkend\@empty\citeyear[#1][#2]{#3}}}
\makeatother

\begin{document}

\title{Lyman continuum leakage in faint star-forming galaxies at redshift $z=3-3.5$ probed by gamma-ray bursts \thanks{Results based on observations carried out at ESO Observatory, Paranal, Chile, by the Stargate consortium under Program ID: 0104.D-0600, P.I.: N. Tanvir }}

\author{
J.-B.~Vielfaure\inst{1},
S.~D.~Vergani\inst{1,2},
J.~Japelj\inst{3},
J.~P.~U.~Fynbo\inst{4,5},
M.~Gronke\inst{6}\thanks{Hubble fellow},
K.~E.~Heintz\inst{7},
D.~B.~Malesani\inst{8},
P.~Petitjean\inst{2},
N.~R.~Tanvir\inst{9},
V.~D'Elia\inst{10,11},
D.~A.~Kann\inst{12}, 
J.~T.~Palmerio\inst{1}, 
R.~Salvaterra\inst{13},
K.~Wiersema\inst{14,15},
M.~Arabsalmani\inst{16,17},
S.~Campana\inst{18},
S.~Covino\inst{18},
M.~ De Pasquale\inst{19},
A.~ de Ugarte Postigo\inst{12},
F.~Hammer\inst{1},
D.~H.~Hartmann\inst{20},
P.~Jakobsson\inst{7},
C.~Kouveliotou\inst{21,22},
T.~Laskar\inst{23},
A.~J.~Levan\inst{15, 24} and
A.~Rossi\inst{25}
}

\institute{\inst{1} GEPI, Observatoire de Paris, PSL University, CNRS, Place Jules Janssen, 92190 Meudon, France \\
\inst{2} Institut d'Astrophysique de Paris, UMR 7095, CNRS-SU, 98 bis boulevard Arago, 75014, Paris, France\\
\inst{3} Anton Pannekoek Institute for Astronomy, University of Amsterdam, Science Park 904, 1098 XH Amsterdam, The Netherlands \\
\inst{4} Cosmic Dawn Center (DAWN)\\
\inst{5} Niels Bohr Institute, University of Copenhagen, Juliane Maries Vej 30, 2100 Copenhagen {\o}, Denmark \\
\inst{6} Department of Physics \& Astronomy, Johns Hopkins University, Baltimore, MD 21218, USA\\
\inst{7} Centre for Astrophysics and Cosmology, Science Institute, University of Iceland, Dunhagi 5, 107 Reykjav\'ik, Iceland \\
\inst{8} DTU Space, National Space Institute, Technical University of Denmark, DK-2800 Kongens Lyngby, Denmark \\
\inst{9} Department of Physics \& Astronomy and Leicester Institute of Space \& Earth Observation, University of Leicester, University
Road, Leicester LE1 7RH, UK \\
\inst{10} ASI-Space Science Data Center, via del Politecnico snc, 00133 Rome, Italy \\
\inst{11} INAF – Osservatorio Astronomico di Roma, via Frascati 33, I-00040 Monteporzio Catone, Italy \\
\inst{12} Instituto de Astrof\'isica de Andaluc\'ia (IAA-CSIC), Glorieta de la Astronom\'ia s/n, 18008 Granada, Spain \\
\inst{13} INAF – IASF/Milano, via Corti 12, I-20133 Milano, Italy \\
\inst{14} Department of Physics and Astronomy, University of Leicester, University Road, Leicester LE1 7RH, UK \\
\inst{15} Department of Physics, University of Warwick, Coventry CV4 7AL, UK \\
\inst{16} IRFU, CEA, Université Paris-Saclay, 91191 Gif-sur-Yvette, France \\
\inst{17} Université Paris Diderot, AIM, Sorbonne Paris Cité, CEA, CNRS, 91191 Gif-sur-Yvette, France \\
\inst{18} INAF - Osservatorio Astronomico di Brera, via Bianchi 46, 23807, Merate (LC), Italy \\
\inst{19} Department of Astronomy and Space Sciences, Istanbul University, Beyaz\i t, 34119, Istanbul, Turkey \\
\inst{20} Department of Physics and Astronomy, Clemson University, Clemson, SC29634-0978, USA \\
\inst{21} Department of Physics, the George Washington University, 725 21st Street NW, Washington, DC 20052, USA \\
\inst{22} Astronomy, Physics and Statistics Institute of Sciences (APSIS), 725 21st Street NW, Washington, DC 20052, USA \\
\inst{23} Department of Physics, University of Bath, Claverton Down, Bath, BA2 7AY, UK \\
\inst{24} Department of Astrophysics/IMAPP, Radboud University, Nijmegen, The Netherlands \\
\inst{25} INAF – Osservatorio di Astrofisica e Scienza dello Spazio, via Piero Gobetti 93/3, 40129 Bologna, Italy \\
}

\date{Received 1 May 2020 / Accepted 14 June 2020 }

  \abstract
  {
The identification of the sources that reionized the Universe and their specific contribution to this process are key missing pieces of our knowledge of the early Universe.
Faint star-forming galaxies may be the main contributors to the ionizing photon budget during the epoch of reionization (EoR), but their escaping photons cannot be detected directly due to inter-galactic medium opacity.
Hence, it is essential to characterize the properties of faint galaxies with significant Lyman continuum (LyC) photon leakage up to $z$\,$\sim$\,4 to define indirect indicators allowing analogs to be found at the highest redshift.}
{Long gamma-ray bursts (LGRBs) typically explode in star-forming regions of faint, star-forming galaxies. Through LGRB afterglow spectroscopy it is possible to detect directly LyC photons. Our aim is to use LGRBs as tools to study LyC leakage from faint, star-forming galaxies at high redshift.}
{Here we present the observations of LyC emission in the afterglow spectra of \object{GRB 191004B} at $z=3.5055$, together with those of the other two previously known LyC-leaking LGRB host galaxies (\object{GRB\,050908} at $z=3.3467$, and \object{GRB\,060607A} at $z=3.0749$), to determine their LyC escape fraction and compare their properties.}
{From the afterglow spectrum of GRB\,191004B we determine a neutral hydrogen 
column density at the LGRB redshift of \lognhicm $~= 17.2 \pm 0.15$, and negligible extinction ($A_{\rm V}=0.03 \pm 0.02$ mag). The only metal absorption lines detected are \ion{C}{iv} and \ion{Si}{iv}.
In contrast to GRB\,050908 and GRB\,060607A, the host galaxy of GRB\,191004B displays significant Lyman-alpha (\Lya) emission. 
From its \Lya~emission and the non-detection of Balmer emission lines we constrain its star-formation rate (SFR) to $1 \leq$ SFR $\leq 4.7$ \Msunyr. We fit the \Lya~emission with a shell model and find parameters values consistent with the observed ones. 
The absolute (relative) LyC escape fractions we find for GRB\,191004B, GRB\,050908 and GRB\,060607A are of $0.35^{+0.10}_{-0.11}$ ($0.43^{+0.12}_{-0.13}$), $0.08^{+0.05}_{-0.04}$ ($0.08^{+0.05}_{-0.04}$) and $0.20^{+0.05}_{-0.05}$ ($0.45^{+0.15}_{-0.15}$), respectively.
We compare the LyC escape fraction of LGRBs to the values of other LyC emitters found from the literature, showing that LGRB afterglows can be powerful tools to study LyC escape for faint high-redshift star-forming galaxies. Indeed we could push LyC leakage studies to much higher absolute magnitudes. The host galaxies of the three LGRBs presented here have all $M_{\rm 1600}>-19.5$ mag, with the GRB\,060607A host at $M_{\rm 1600}>-16$ mag. LGRB hosts may therefore be particularly suitable for exploring the ionizing escape fraction in galaxies that are too faint or distant for conventional techniques.
Furthermore, the time involved is minimal compared to galaxy studies.} 
  {}

   \keywords{galaxies: high-redshift -- intergalactic medium -- gamma-ray burst: general -- dark ages, reionization, first stars -- galaxies: evolution}

\titlerunning{LyC leakage in LGRB hosts}
\authorrunning{J-B. Vielfaure et al.}

   \maketitle

\section{Introduction}
Understanding the nature of the sources responsible for ionizing hydrogen in the intergalactic medium (IGM) remains one of the key challenges in studies of early structure formation.
It has been established that active galactic nuclei on their own provide enough ionizing radiation to keep the Universe reionized at redshifts $z < 4$ \citep[e.g.,][]{Cristiani2016}. According to the predominant view, star-forming galaxies (SFGs) are the main contributors of ionizing radiation at earlier epochs \citep[$z \gtrsim 4$; e.g.,][]{Fontanot2012,Robertson2013,Robertson2015}. 

In order to quantify the contribution of SFGs to the reionization at $z \gtrsim 4$, we have to know: {\it (i)} the galaxy number density as a function of redshift and luminosity; {\it(ii)} their ionizing photon production efficiency (i.e., the number of hydrogen-ionizing photons relative to produced UV photons at $\sim$1500~\AA, $(f_{1500}/f_{900})^{\rm int}$); {\it(iii)} the fraction of produced Lyman continuum (LyC) photons that can actually escape the local environment and ionize the IGM, called the escape fraction $f_{\rm esc}$. Especially the latter is poorly constrained from observations. 

Direct searches for LyC emission from galaxies is only possible up to $z$\,$\sim$\,4, beyond which the LyC is unobservable due to the increasing IGM opacity \citep[e.g.,][]{Madau1995}. We therefore rely on lower redshift galaxies to be used as proxies for the higher-redshift ones. The search for LyC emitters has been quite fruitful in recent years. A few LyC emitters have been found in the local Universe \citep{Bergvall2006,Leitet2013,Borthakur2014}. A population of green pea galaxies with strong LyC has been identified at $z$\,$\sim$\,0.3 \citep{Izotov2016b,Izotov2018a,Izotov2018b}. A number of LyC-emitting galaxies with typically $M_{\rm UV} < -19.5$ mag have also been securely identified at higher ($z \gtrsim 3$) redshifts \citep{Vanzella2016,Vanzella2018,Shapley2016,Bian2017,Steidel2018,Fletcher2019}. The growing number of discovered LyC emitters allows us to look at their common properties that could be used both to make it easier to find new emitters at these redshifts as well as to identify likely emitters at $z > 6$. The emitters are typically (but not always) found to have a strong Ly$\alpha$ emission line \citep{Verhamme2015,Verhamme2017}, a high ratio of [OIII]/[OII] nebular emission lines \citep{Nakajima2014,Nakajima2016} and a compact morphology \citep{Izotov2016b,Izotov2018a,Izotov2018b}. 
The compactness of the galaxy as a requirement for LyC photons to escape is also found in theoretical studies \citet[e.g.,][]{Cen2020}, together with high star-formation rate surface densities.

Recently, important constraints on the LyC emission at high redshift come from an analysis of stacked spectra of galaxies at $-19.5 > M_{\rm UV} > -22$ mag at $z$\,$\sim$\,3 \citep{Steidel2018}. 
The measured mean escape fraction of $0.09\pm0.01$ is dominated by sub-L$_{\rm UV}$* galaxies, with negligible contribution from L > L$_{\rm UV}$* systems.
These results are in agreement with the hypothesis that fainter galaxies have on average higher escape fractions and account for a large part of the ionizing-photon budget \citep[e.g.,][]{Fontanot2014}. 
Because the bulk of the high-redshift star formation occurred in very faint galaxies \citep[e.g.,][]{Bouwens2015}, the regime of LyC emission in those galaxies needs to be studied observationally \citep[see, e.g.,][]{Japelj2017}. Deep observations of galaxies in this faint regime ($M_{\rm UV} \gtrsim -19$ mag) in the LyC portion of their spectra are time-demanding and therefore sparse. \citet{Fletcher2019} identified four faint LyC emission candidates and \citet{Amorin2014b} studied a faint lensed galaxy resulting in a poorly constrained limit on its LyC escape fraction. 

A unique way to learn more about the escape fractions of faint galaxies is to make use of long gamma-ray bursts (LGRBs). LGRBs mark the deaths of massive stars \citep{Hjorth2003a,Woosley2006a,Cano2017} and can be observed from the local to the $z$\,$\sim$\,8 Universe \citep{Tanvir2009,Salvaterra2009}. LGRBs are generally found in sub-L* galaxies at any redshift \citep[e.g.,][]{Tanvir2012,Vergani2015,Perley2016b,Palmerio2019}. The progenitors are associated with young star-forming regions that contribute hugely to the ionizing photon budget inside galaxies \citep[e.g.,][]{Ramachandran2018}. A burst of gamma-ray emission is typically followed by a longer-lived, multiwavelength afterglow
\citep[e.g.,][]{Kumar2015}. Thanks to their brightness, the spectra of the UV/optical/NIR afterglows can reveal the detailed properties of the interstellar medium (ISM) along the lines of sight toward star-forming regions, such as hydrogen column density, dust properties, metallicity and kinematics \citep[e.g.,][]{DePasquale2003, Thone2014, Heintz2018, Zafar2018a, Bolmer2019}. In addition, the afterglow brightness allows us to search directly for possible leaking LyC in their spectra. With LGRB afterglows we can therefore study the transparency of the LGRB star-forming regions to ionizing radiation \citep{Chen2007b,Fynbo2009,Tanvir2019}. 
Being based on afterglow spectroscopy, LGRB LyC studies do not suffer from galaxy magnitude selection and offer the advantage of a direct determination of the $(f_{1500}/f_{900})^{\rm int}$ ratio thanks to the
intrinsically featureless (power-law) spectra of GRB afterglows.
LGRB afterglows are thus a very useful tool to search for LyC leakage in very faint galaxies at $z$\,$\sim$\,$2-4$. 

In this paper we present the detection of LyC photons in the afterglow spectrum of GRB\,191004B at $z = 3.5055$. We further investigate LyC emission in LGRB afterglow spectra, also adding to this case GRBs 050908 and 060607A ($z=3.3467$ and $z=3.0749$, respectively), the only two previously known GRBs whose spectra show  significant LyC emission \citep{Fynbo2009,Tanvir2019}. This paper is organized as follows. In Section 2 we present the observations of GRB\,191004B. We analyze them in Section 3. In Sections 4 and 5 we determine the escape fraction along the lines of sight to the three GRBs. Section 6 is focused on the \Lya~emission modeling. All the results are discussed in Section 7, especially in the context of other known LyC-emitting galaxies. We provide our conclusions in Section 8.

All errors are reported at 1$\sigma$ confidence and all magnitudes are reported in AB unless stated otherwise. We consider a flat $\Lambda CDM$ cosmology with the cosmological parameters provided in \citet{Planck2016}: $ H_0 = 67.8$ km s$^{-1}$ Mpc$^{-1}$, $ \Omega_{\rm m} = 0.308$ and $ \Omega_{\Lambda} = 0.692$.

\section{Observations}

GRB\,191004B was detected by the BAT instrument on the \textit{Neil Gehrels Swift Observatory} (henceforth \textit{Swift}) on 2019 October 4, at 21:33:41 UT (Trigger 927839; \citealt{Cenko2019GCN}).
The BAT light curve shows a multi-peaked shape with a duration of $t_{90} (15-350$ keV) $= 37.7\pm14.9$~s \citep{Stamatikos2019GCN}.
The {\it Swift} UV/Optical Telescope started settled observations of GRB\,191004B 69~s after the trigger and refined the position to RA(J2000)$=03^{\rm h} 16^{\rm m} 49.10^{\rm s}$, Dec(J2000)$=\ang{-39} 38' 03\farc9$ with a 90\% confidence error radius of $\sim$\,0\farc4 \citep{LaPorte2019GCN}. The redshift of $z = 3.503$ was reported by \citet{DElia2019GCN} based on VLT/X-shooter observations of absorption lines in the optical afterglow (see also below).

In the following we describe our ESO/VLT observations of the GRB\,191004B afterglow and host galaxy obtained by the {\it Stargate} consortium under Program ID: 0104.D-0600 (P.I.: N. Tanvir). The log of observations is provided in Table \ref{Tab:observations}.

\subsection{VLT/X-shooter afterglow imaging}
\label{Obs_OA}

GRB\,191004B was observed with the VLT/X-shooter acquisition camera in the $g^\prime$, $r^\prime$, $z^\prime$ bands $\sim$7.2 hours after the burst. The observations were taken in a sequence of $3 \times 40$s, $3 \times 30$s and $3 \times 60$s with the $g^\prime$, $r^\prime$ and $z^\prime$ filters, respectively. 
The afterglow is clearly detected in all bands at the coordinates RA(J2000) $= 03^{\rm h}16^{\rm m}49^{\rm s}.14$, Dec(J2000) $= \ang{-39}38'$04\farc09, with a magnitude of $g^\prime$ = $22.41 \pm 0.03$ mag, $r^\prime$ = $21.23 \pm 0.02$ mag and $z^\prime$ = $20.91 \pm 0.04$ mag.
The photometry was calibrated with a single nearby star due to the small field of view. The coordinates of this star RA(J2000)$=03^{\rm h} 16^{\rm m} 54^{\rm s}.42$, Dec(J2000)$=-39^{\rm d} 37' 41\farc9$ are taken from the Gaia catalog and the photometry from the Pan-STARRS catalog.

\subsection{VLT/X-shooter afterglow spectroscopy } \label{xshOA}

The first epoch spectrum of GRB\,191004B was obtained with the VLT/X-shooter echelle spectrograph \citep{Vernet2011} just after the $g^\prime$, $r^\prime$, $z^\prime$ imaging, $\sim$7.2 hours after the {\it Swift}/BAT trigger. The observation consisted of a total integration time of 2400~s in each arm using the 1\farc0/0\farc9/0\farc9JH slits at a position angle of PA$=\ang{-75}$ (see Fig.~\ref{FORS2_image}).
The observation was carried out under good conditions with an average seeing of $\sim$ 0\farc75 and an airmass of 1.1.
We used a nodding scheme with a 5\farc0 nodding throw. 
This observation covers the afterglow emission of the GRB across the wavelength range 300 - 2100~nm. 

We reduced the X-shooter data using version 3.3.5 of the X-shooter data reduction pipeline \citep{Modigliani2010}.
All observations of the afterglow and host galaxy, the telluric stars, and the spectrophotometric standards were reduced in the same way. Before processing the spectra through the pipeline, the cosmic-ray hits and bad pixels were removed following the method of \citet{VanDokkum2001}.
Then, we subtracted the bias from all raw frames and divided them by the master flat field. We traced the echelle orders and calibrated the data in spatial and wavelength units using arc-line lamps. The flux calibration was done using spectrophotometric standards \citep{Vernet2009} and a correction for mirror flexure was applied. Lastly, the sky-subtraction and the rectification and merging of the orders was done to obtain the final two-dimensional spectra.
To optimally select the extraction region we chose the spatial extension of the afterglow continuum
independently in each of the three (UVB, VIS, NIR) arms of the spectrograph.
We corrected the 1D spectra for Galactic extinction using the Milky-Way extinction curve of \citet{Pei1992} (assuming the ratio of total-to-selective extinction $R_V=$ 3.08) and the extinction map of \citet{Schlafly2011}.

\begin{table}
\begin{center}
\caption{Log of observations for the GRB\,191004B afterglow and host galaxy.}
\label{Tab:observations}
\small
\begin{tabular}{ c c c c c c } \hline \hline
          Instrument      & $\Delta t$    & $T_{\rm exp}$ & Slits / Filters & Seeing \T \\
                          & (day)   & (s)         &                &   (")           \B \\
          \hline
          X-shooter   & 0.30 & 2400/2400/2400 & 1.0/0.9/0.9JH & 0.75 \T\B \\
          X-shooter   & 108  & 5840/5440/6000 & 1.6/1.5/0.9 & 0.6  \B \\
         \hline
		  X-shooter   & 0.30 & 3x40/3x30/3x60 & $g^\prime$, $r^\prime$, $z^\prime$   & 0.75 \T\B \\
          FORS2       & 77   & 1800           & $R\_SPECIAL$           & 0.7 \B \\
  \hline
\end{tabular}
\tablefoot{The columns indicate the instrument, the time $\Delta t$ after the GRB detection, the total exposure time $T_{\rm exp}$ in each slit/filter, the slit/filter configuration in the case of spectral/imaging observations and the average DIMM seeing at the time of the observation.}
\end{center}
\end{table}

This final data set typically has a signal to noise ratio S/N > 3 per spectral bin and 
a spectral resolution of $R$\,$\sim$\,6700/11000/7800 in the UVB/VIS/NIR arms, respectively.
The resolution was estimated from the average seeing reached during the observation corrected for airmass and wavelength dependence and by interpolating the nominal values available on the ESO instrument website\footnote{\url{https://www.eso.org/sci/facilities/paranal/instruments/xshooter/inst.html}}. 
We also compare these values to the resolutions obtained using the relation derived in \citet{Selsing2019} and we find a good agreement between both.  
We clearly detect the afterglow continuum from 3675~\AA~to the end of the X-shooter spectrum. We identify the \Lya~emission and absorption line at $\sim$5474~\AA~and detect the metal absorption lines described in detail in Sect.~\ref{absorptions}.

\subsection{Host-galaxy imaging and spectroscopy} \label{Obs_Host}
        
\begin{figure}
\centering
\includegraphics[width=\hsize]{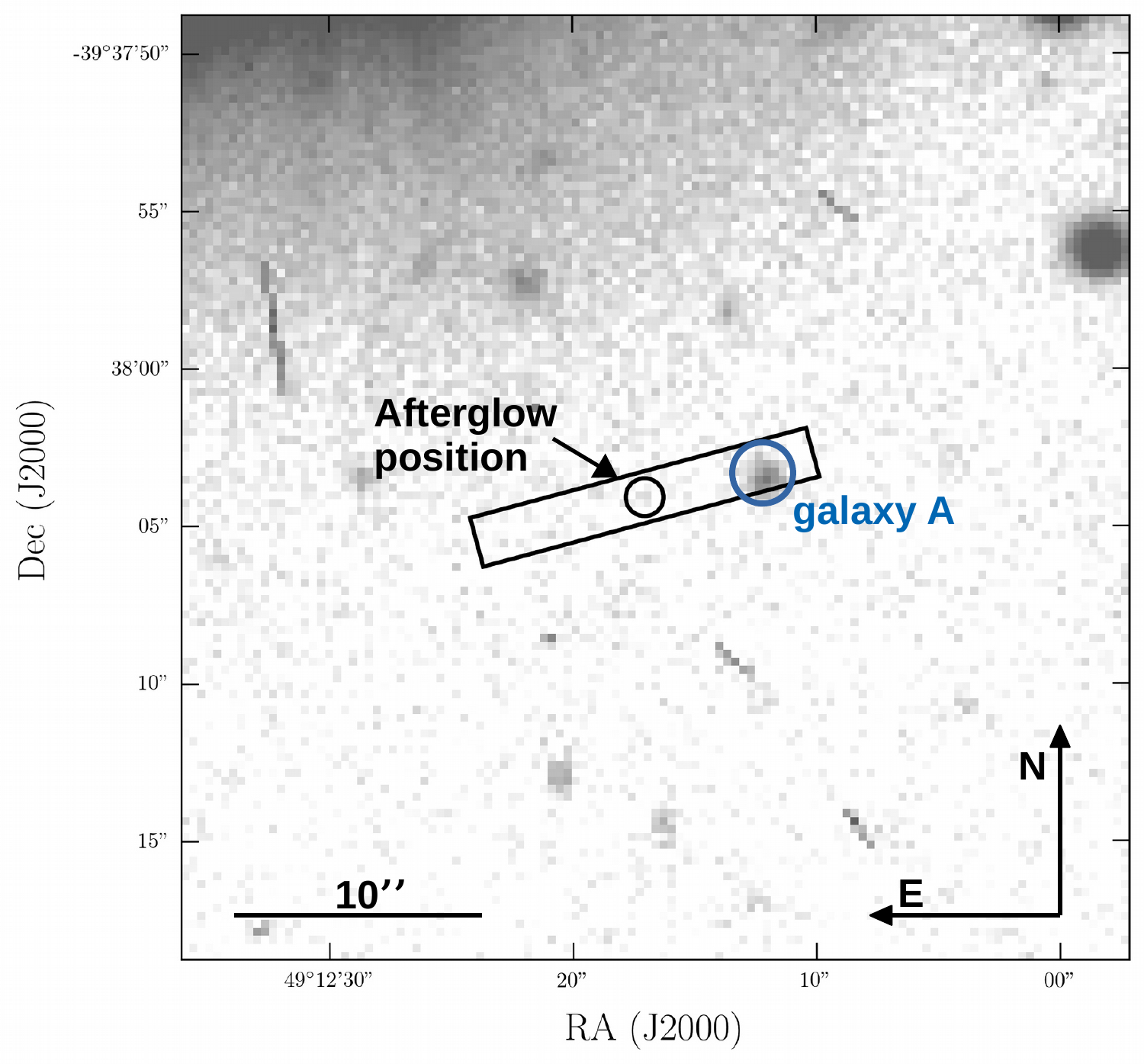}
 \caption{VLT/FORS2 image of the GRB\,191004B host-galaxy field $\sim$77.2 days after the {\it Swift}/BAT trigger. The host galaxy is not detected (magnitude limit {\it R} $\geq 26.6$~mag). The corresponding afterglow position (error radius $r_e=0\farc6$, as observed in the X-shooter $r^\prime$ image $\sim$7.2 hours after the burst) is indicated by a black circle. The slit at PA$=\ang{-75}$ used for the X-shooter afterglow spectroscopy is also overplotted.
 Galaxy $A$ corresponds to the intervening system at $z=2.137$ identified in the VLT/X-shooter spectrum.  
 }
 \label{FORS2_image}
\end{figure}

We observed the field of GRB\,191004B with the VLT/FORS2 \citep{Appenzeller1998} camera in imaging mode on 2019 December 21, at 03:25:03 UT ($\sim$77 days after the {\it Swift}/BAT trigger).  
In the 30~min {\it R}-band image (see Fig.~\ref{FORS2_image}), we do not detect the host galaxy. The magnitude limit is $R \geq 26.6$~mag.

A further VLT/X-shooter spectrum was taken at the afterglow position, once the afterglow faded ($\sim$108 days after the {\it Swift}/BAT trigger), using the blind offset technique to look for host galaxy emission lines. The average seeing and airmass during the observation were $\sim$0\farc6 and 1.2, respectively.
The observations were divided in two OBs for a total integration time of 5840/5440/6000 s in the UVB/VIS/NIR arm, respectively. We used the 1\farc6/1\farc5/0\farc9 slits at a position angle of $\ang{-40}$ and a nodding scheme with a 5\farc0 nodding throw. The data reduction was carried out in a similar fashion as described in Sect.~\ref{Obs_OA}. 
We corrected the 1D spectra for Galactic extinction as for the afterglow spectrum (Sect.~\ref{xshOA}). 
We clearly detect the \Lya~emission line at the same wavelength as in the afterglow spectrum but we do not detect other nebular emission lines. 
Unlike the afterglow spectra, no continuum is detected in the host-galaxy spectra, so to optimally select the extraction region we chose the spatial extension of the Ly$\alpha$ emission line and applied this 1D extraction throughout the whole spectrum.
We provide flux and upper limits of the lines in Sect.~\ref{absorptions}.
The good seeing obtained during this observation allows us to reach resolutions of $R$\,$\sim$\,7800/12900/9200 in the UVB/VIS/NIR arms, respectively, calculated similarly as in Sect.~\ref{xshOA}.
    
\section{Data analysis} \label{results}

\subsection{\HI~and metal absorption lines} \label{absorptions}

\begin{figure}
\centering
\includegraphics[width=0.90\hsize]{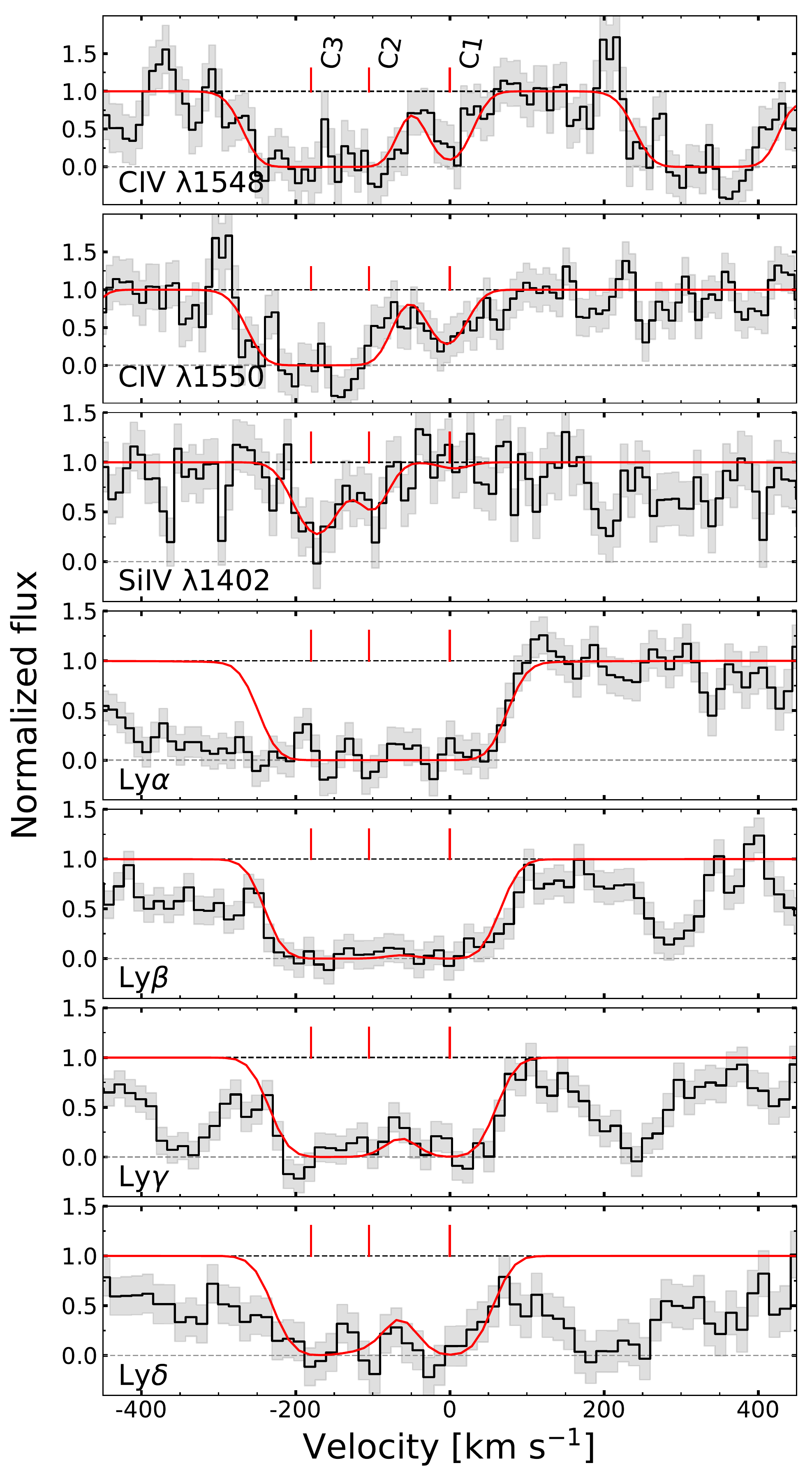}
 \caption{\ion{C}{iv}, \ion{Si}{iv} and Lyman series absorptions of the GRB\,191004B host galaxy identified in the VLT/X-shooter afterglow spectrum. The gray regions correspond to the uncertainty on the flux and the red line to the best-fit model, obtained using three components. 
Zero corresponds to the reddest component at $z=3.5055$ named $C1$. $C2$ and $C3$ are the two other components detected at -105 and -180 \kms, respectively.
}
 \label{fig:fit_abs_GRB191004B}
 \label{aftfit}
\end{figure}

The X-shooter afterglow spectrum reveals Ly$\alpha$, Ly$\beta$, Ly$\gamma$, and Ly$\delta$, \ion{Si}{iv} and \ion{C}{iv} absorption lines from the host galaxy gas \citep{DElia2019GCN} (see Fig.~\ref{aftfit}).

Firstly, we fit the saturated \ion{Si}{iv} and \ion{C}{iv} absorption with three components using \texttt{VoigtFit} \citep{Krogager2018}. We determine lower limits on the \ion{C}{iv} and \ion{Si}{iv} column densities of $\log(N_{\rm X}$/cm$^{-2})> 15.6$ and 14.1, respectively, and a redshift of $z=3.5028$, 3.5039, and 3.5055 for the three components (corresponding to relative velocity from the reddest component of $\sim$ -105 and -180 \kms, respectively). The Doppler parameters are $b=25$, 20, 30 \kms~(from the red component to the blue one, respectively). 

The high-ionization metal lines such as \ion{C}{iv} and \ion{Si}{iv} do not necessarily originate from the same place as the neutral gas \citep[e.g.,][]{Heintz2018}. However, since they are the only detected for the GRB system, we fixed these components and their redshift to fit the Lyman lines. The Lyman lines are saturated, therefore their fitting could not give a precise estimate of the hydrogen column density. 
Fixing the $b$ parameters to those inferred from the corresponding metal line components,
we find an acceptable fit for \lognhicm~$=17.2\pm0.3$. A fit letting the redshift of the components and the $b$ parameters free to vary would give consistent \NHI~values. This is one of the lowest hydrogen column densities measured among the LGRB afterglow sample \citep{Tanvir2019} and places the absorbing system on the boundary between being a \Lya~forest absorber (\lognhicm~$<17$; \citealt{Rauch98}) 
or a Lyman-limit system (LLS, $17<$ \lognhicm~$< 20.3$; \citealt{Peroux2003b}). 
The reddest component is the richest in neutral hydrogen, whereas the bluest component is the strongest for the high-ionization lines.

We do not detect the other common singly ionized metal lines such as \ion{Fe}{II}, \ion{Mg}{II}, \ion{Al}{II} or \ion{Si}{II} \citep{Christensen2011,DeUgartePostigo2012}, that are typically detected in LGRB afterglow spectra. 
The part of the spectrum at the corresponding wavelengths of most of the low-ionization lines is quite noisy or contaminated by residual sky lines. We place approximate upper limits on their rest-frame equivalent widths of $\sim$0.4~\AA. We can determine a 3$\sigma$ upper limit on the \ion{C}{II} column density of \lognhicm~$<~14.7$.
Low column densities of low-ionization metal transitions are also found for a part of the LLS sample detected through quasar spectroscopy (see, e.g., \citealt{Fox2013,Cooper2015}), as well as in the other few LGRBs having \lognhicm~$<~19$ \citep[see, e.g.,][]{Thone2011}. 
In these systems a significant fraction of the gas is ionized, whereas usually LGRB hosts show higher \HI~absorption values (median of \lognhicm~$=~21.6$; \citealt{Tanvir2019}), typical of systems probing neutral gas. At such low \HI~column densities, even for approximately solar metallicity gas, the strongest low-ionization metal absorption features can have column densities below the detection threshold. 
\cite{DeCia2011} found very low values for the column densities of low-ionization metals detected in GRB\,070125 afterglow spectra, at $z=1.5477$. They conclude that the LGRB was likely located in the outskirts of a massive star-forming region inside a faint and small host galaxy \citep[see also ][]{Updike2008, Cenko2008}.

We detect \ion{Fe}{II}, \ion{Mg}{II} absorption lines and [\ion{O}{III}] emission corresponding to an intervening absorber at $z=2.137$. This system is at 3\farc8 from the afterglow position and corresponds to galaxy A in Fig.~\ref{FORS2_image}. Its emission trace in the 2D spectrum covers 1\farc5. 
We do not detect the presence of another galaxy brighter than $R$=26.6~mag in a radius of 3\farc0 around the position of the GRB\,191004B afterglow, corresponding to 22 kpc at this redshift. 
Therefore, we can rule out contamination of the observed LyC emission from foreground interlopers.

\subsection{Lyman continuum leakage} \label{LyC}
    
\begin{figure*}[!ht]
\centering
\includegraphics[width=0.75\textwidth]{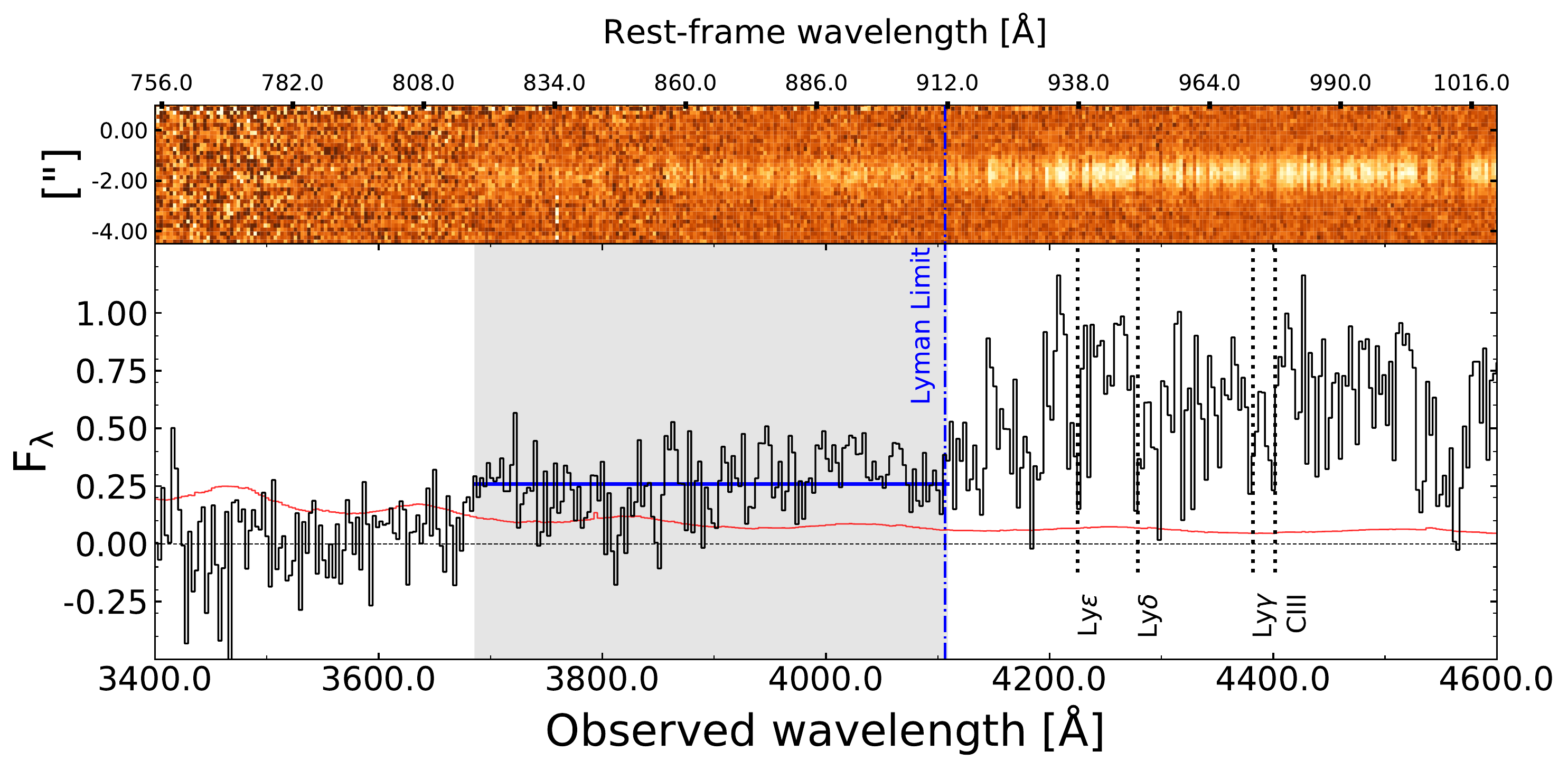}
 \caption{Part of the afterglow UVB spectrum of the GRB\,191004B covering the 912~\AA~region (rest-frame). Top: 2D spectrum showing the detection of residual emission below the Lyman limit.
 Bottom: 1D spectrum (black line) with flux density $F_{\lambda}$ in units of $ 10^{-17}$ \ergscmA. Its associated error spectrum (1$\sigma$) is in red. The grey shaded region marks the rest-frame range 816 - 910~\AA~where residual flux below 912~\AA~is detected.
 The solid blue line represents the corresponding average LyC flux density of $f_{900} = (2.6 \pm 0.2) \times 10^{-18}$ \ergscmA over this range.
The spectrum is rebinned at 3~\AA~per pixel.}
 \label{GRB191004B_2D_1D}
\end{figure*}

The afterglow spectrum of GRB\,191004B clearly presents residual flux in the observed wavelength range 3675.00-4106.73~\AA~(816 - 912~\AA~rest-frame), see Fig.~\ref{GRB191004B_2D_1D}. 
We measure an average flux density of $f_{900}^{\rm obs} = (2.6 \pm 0.2) \times 10^{-18}$ \ergscmA~over this wavelength range. 

We detect several \Lya~absorbers in the \Lya~forest from $z=2.8$ to $z=3.45$, with hydrogen column densities spanning log(\NHI$/\rm cm^{-2})\sim15-17$. The lack of LyC emission below $\sim$3650~\AA~is due to the Lyman limit absorption by some of these systems, and the dimming of the LyC emission around 3800~\AA~corresponds to the Lyman limit of a system at $z=3.19$ and to the likely strong \Lya~absorption of Galaxy A (see Sect.~\ref{absorptions}).

The non-ionizing UV flux density measured in the rest-frame range 1480 - 1520~\AA~is $ f_{1500}^{\rm obs} = (5.5 \pm 0.1) \times 10^{-18}$ \ergscmA. Together they correspond to a non-ionizing to ionizing flux ratio of $(f_{1500} / f_{900})^{\rm obs} = 2.1 \pm 0.2$. This ratio will be used to measure the escape fraction of ionizing radiation in Sect.~\ref{DirectMeasurement}.

The host-galaxy spectrum does not show LyC emission above a 3$\sigma$ flux density limit of $4.5 \times 10^{-18}$ \ergscmA. 
A rough binning of the spectrum in the 820 - 910~\AA~range allows to place a more stringent 3$\sigma$ upper limit on its LyC flux density to $6.0 \times 10^{-19}$ \ergscmA.

\subsection{Line-of-sight extinction} \label{extinction}

\begin{figure}
\centering
\includegraphics[width=\hsize]{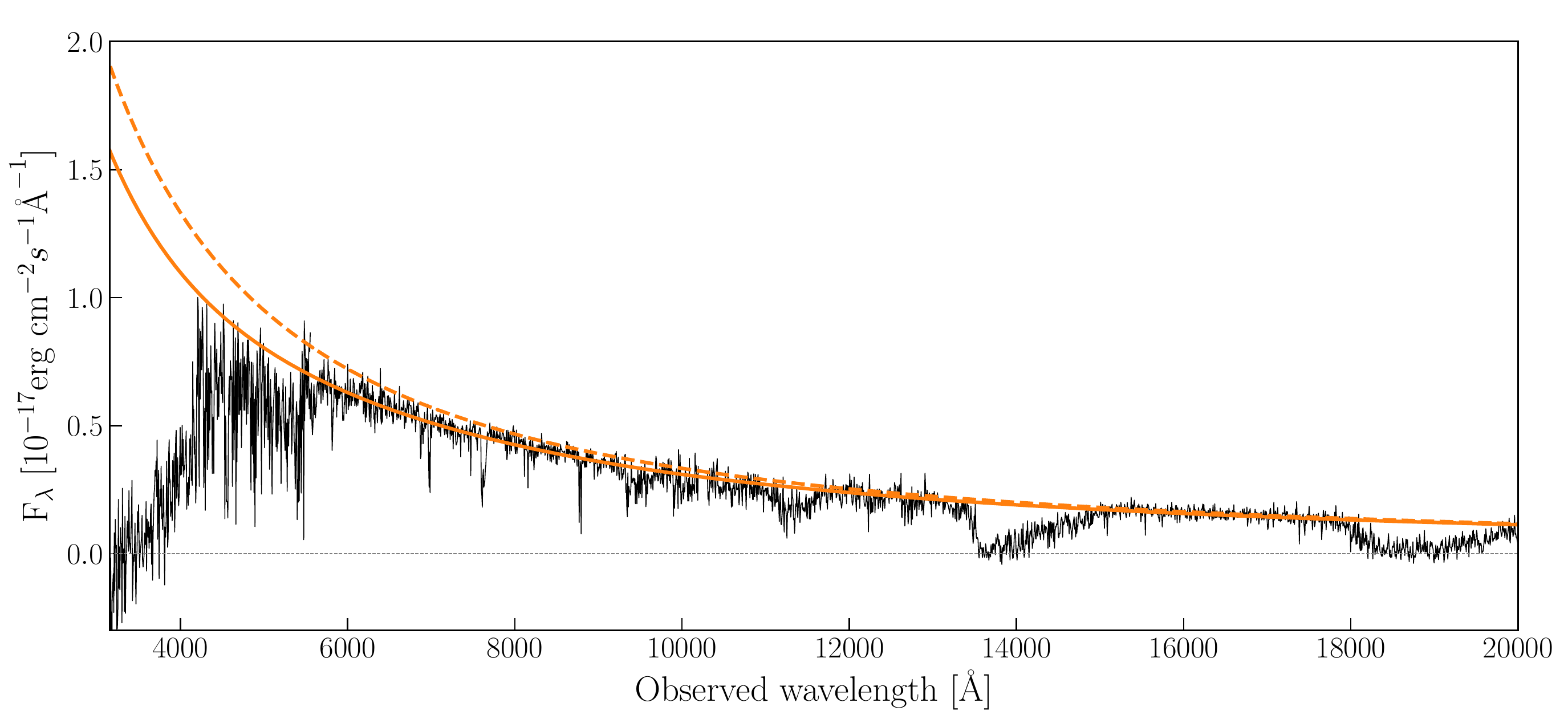}
 \caption{Fit of the dust extinction to the afterglow spectrum of GRB\,191004B (black line).
 The solid orange line corresponds to the best-fit extinction (solid orange line) with a spectral index $\beta_{o}=0.49\pm0.03$ and an SMC extinction law with $A_{V}=0.03^{+0.02}_{-0.02}$ mag. See Sect.~\ref{extinction} for more details. The orange dashed line is for the intrinsic spectrum corrected for extinction.}
 \label{GRB191004B_powerLaw}
\end{figure}

We measured the host-galaxy extinction along the line of sight to the GRB directly from the afterglow spectrum corrected for Galactic extinction (see Sect, \ref{xshOA}) following the procedure as described in \citet{Japelj2015} and \citet{Zafar2018a}.
Due to the sparse X-ray afterglow observations around the epoch of the X-shooter spectrum, we did not include X-ray data into this fit. 
We ignored the spectral region blueward of the \Lya~emission feature in the fit, since it is significantly affected by \Lya~forest absorption. The edges of the VIS and NIR spectra and the regions affected by telluric absorption were also masked out.
We cleaned the spectra of absorption lines by recursively fitting a polynomial to the continuum and removing everything that deviate from the fit by more than $3.5\sigma$.  
Following the procedure of \citet{Japelj2015}, we fitted the spectrum with a power-law function modulated by dust extinction assuming the Milky Way and Magellanic Clouds extinction curves from \citet{Pei1992}. We found that the spectrum can be well described by a power-law with small extinction.
The best fit gives an optical spectral index $\beta_{\rm o} = 0.49 \pm 0.03$, an extinction $A_{\rm V} = 0.03 \pm0.02$~mag and $\chi^{2}/{\rm dof} =  1.007$ 
for the Small Magellanic Cloud extinction curve. We present the best fit (solid orange line) together with the corresponding intrinsic power-law corrected for extinction (dashed line) in Fig.~\ref{GRB191004B_powerLaw}.

\subsection{\Lya~emission} \label{LyaE}

\begin{figure}
\centering
\includegraphics[width=\hsize]{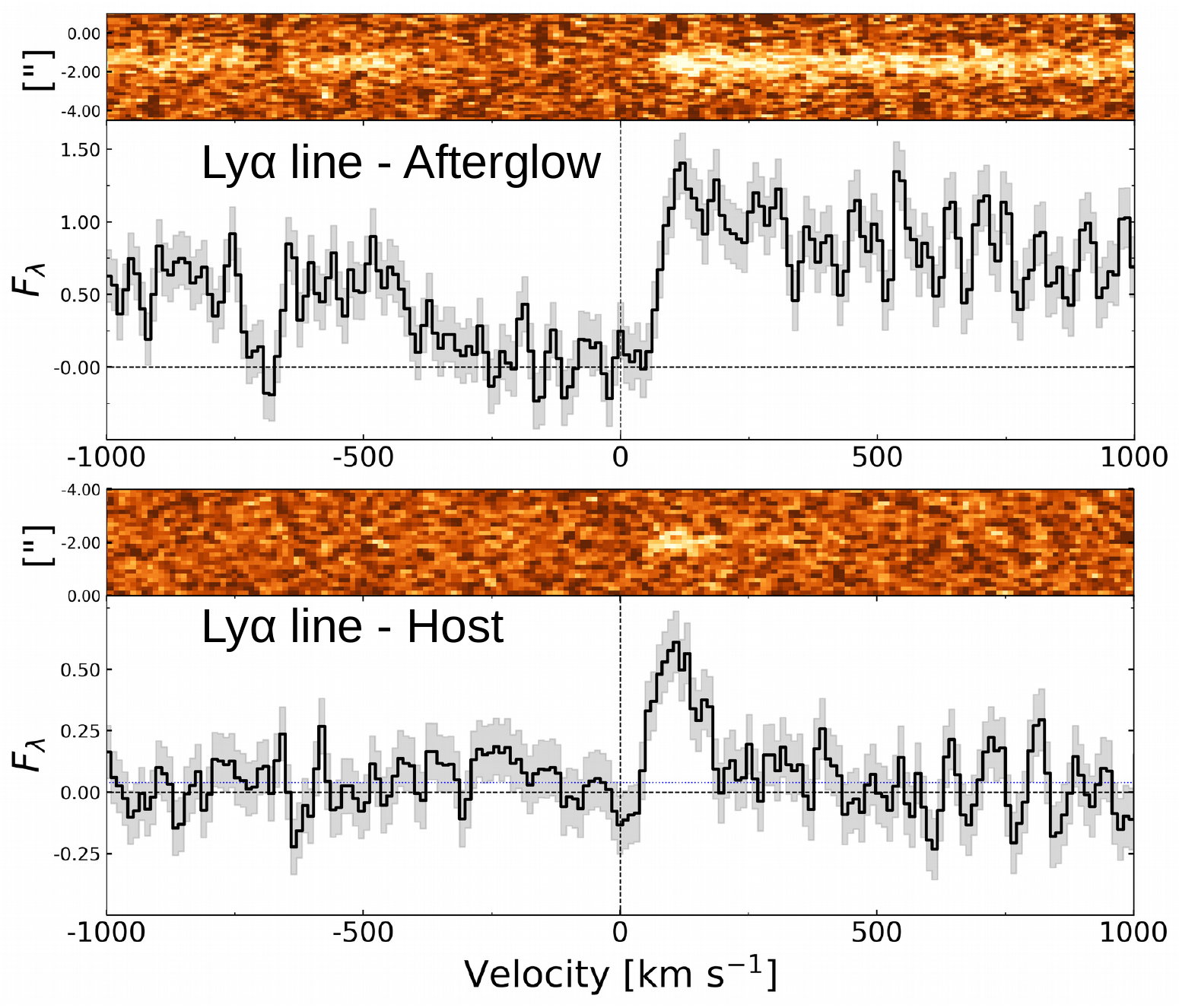}
 \caption{Part of the 2D and 1D UVB spectra of the GRB\,191004B in velocity space, centered on the \Lya~line at $z=3.5055$. 
 The flux density $F_{\lambda}$ is in units of $ 10^{-17}$ \ergscmA. 
 Top: VLT/X-shooter spectrum of the afterglow $\sim$7.2 hours after the burst detection. 
 The \Lya~line is visible in the red wing of the absorption line. 
 Bottom: VLT/X-shooter spectrum of the host galaxy. The \Lya~emission line is clearly detected at $ 113 \pm 30$ \kms.
 }
 \label{Lyman-a}
\end{figure}

We detect the \Lya~emission line in both the X-shooter afterglow and host-galaxy spectra (see Fig.~\ref{Lyman-a}). 
The emission line is asymmetric.
We fit its profile with a skewed Gaussian parametrized as in \citet{Claeyssens2019, Matthee2019},
\begin{equation}
 f(\lambda) = A\ \exp \Bigg[ - \frac{ (\lambda - \lambda_{0})^2 }{ 2\ (a_{\rm asym}\ (\lambda - \lambda_{0}) + d )^2  } \Bigg] \, ,
 \label{SkewedGaussian}
\end{equation} 

\noindent where A is the amplitude of the \Lya~line; $\lambda_{0}$ is its peak wavelength; $a_{\rm asym}$ is the measure of asymmetry of the line and {\it d} the parameter that controls the line width.
These parameters are left free during the fitting with broad uniform priors. 
In order to determine both a robust fit and its associated uncertainty, we use the Markov Chain Monte Carlo sampler \texttt{emcee} \citep{Foreman2013} to maximize the likelihood function using Eq.~(\ref{SkewedGaussian}).
The best fit of the X-shooter \Lya~line detected in the host spectrum provides a peak position of
$\lambda_{0} = 5478.85^{+0.13}_{-0.12}$~\AA, an asymmetry of 
$a_{\rm asym} = 0.33^{+0.05}_{-0.07}$ and a width parameter of
$d = 0.6985^{+0.110}_{-0.103}$. We convert the latter into FWHM according to the analytic expression:
\begin{equation}
 FWHM = \frac{ 2\ \sqrt{2\ \log(2)}\ d }{1-2\ \log(2)\ a_{\rm asym}^2} \, .
 \label{FWHM}
\end{equation} 

\noindent The corresponding rest-frame FWHM of the line is $FWHM_0 = 104 \pm 30$~\kms.
The flux of the \Lya~line corrected for Galactic extinction is $F(\rm Ly\alpha) = (1.0 \pm 0.1) \times 10^{-17}$ \ergscm. This corresponds to a \Lya~luminosity of $L_{\rm Ly\alpha}= (1.18 \pm 0.12)\ \times 10^{42}$ erg~s$^{-1}$ and a rest-frame equivalent width of $EW_0(\rm Ly\alpha) = 7.4 \pm 2.6$~\AA. 

We convert this \Lya~flux to a star formation rate of SFR(\Lya) $\approx 1$ \Msunyr~by using the recombination factor 8.7 from the case B of the theory of recombination \citep{Brocklehurst1971} and the relation from \citet{Kennicutt1998}, scaled to the \citet{Chabrier2003} initial mass function. Nevertheless, we know that this SFR can be largely underestimated due to the complex scattering and destruction process of \Lya~photons in neutral gas and dust. Therefore the SFR derived here only represents a lower limit.

\subsection{Other emission lines} \label{Emissions}

We do not detect other emission lines at the afterglow position in the afterglow and host-galaxy spectra. 
We determine 3$\sigma$ upper limits of $0.3 \times 10^{-17}$ \ergscm~for \Hb~and [OII]$_{\lambda3727}$, and $1.0 \times 10^{-17}$ \ergscmA~for [OIII]$_{\lambda5007}$. 
By assuming that the negligible dust extinction in the line of sight is also valid for the integrated host-galaxy dust content, the \Hb~upper limit can be converted to an SFR $< 4.7$ \Msunyr~considering the intrinsic Balmer decrement \Ha/\Hb~$= 2.86$ \citep{Osterbrock1989} and using the relation from \citet{Kennicutt1998}, scaled to the \citet{Chabrier2003} initial mass function (IMF). This value is consistent with the lower limit derived from the \Lya~line flux and allows us to put an additional constraint on the SFR. This implies an SFR of a few \Msunyr~and an escape fraction of \Lya~photons $f_{\rm esc}$(\Lya) > 0.13.

\section{LyC escape fraction} \label{fesc}

\begin{figure}
\centering
\includegraphics[width=\hsize]{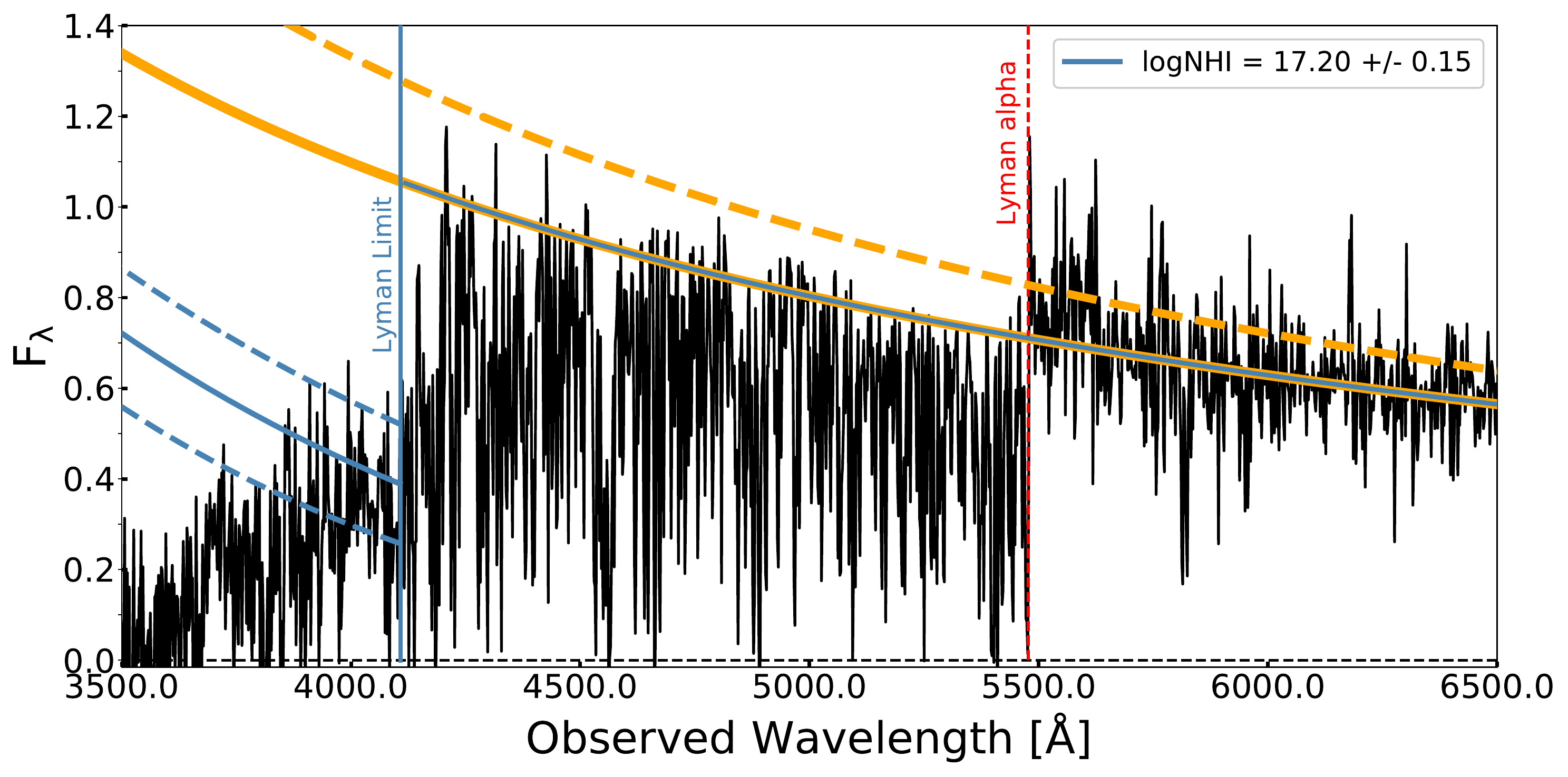}
 \caption{Afterglow spectrum of the GRB\,191004B covering the LL and \Lya~line.
 The flux of the escaping LyC photons obtained for log(\NHI$/\rm cm^{-2})=17.20\pm0.15$ corresponds to the blue solid and dashed lines below the LL. 
 The orange solid (dashed) line corresponds to the fit of the afterglow continuum without (with) correction for dust extinction. 
 The flux density $F_{\rm \lambda}$ is in units of $ 10^{-17}$ \ergscmA.
 The spectrum is rebinned to 2~\AA~per pixel.}
 \label{figFynbo}
\end{figure}

\begin{figure*}
\centering
\includegraphics[width=\textwidth]{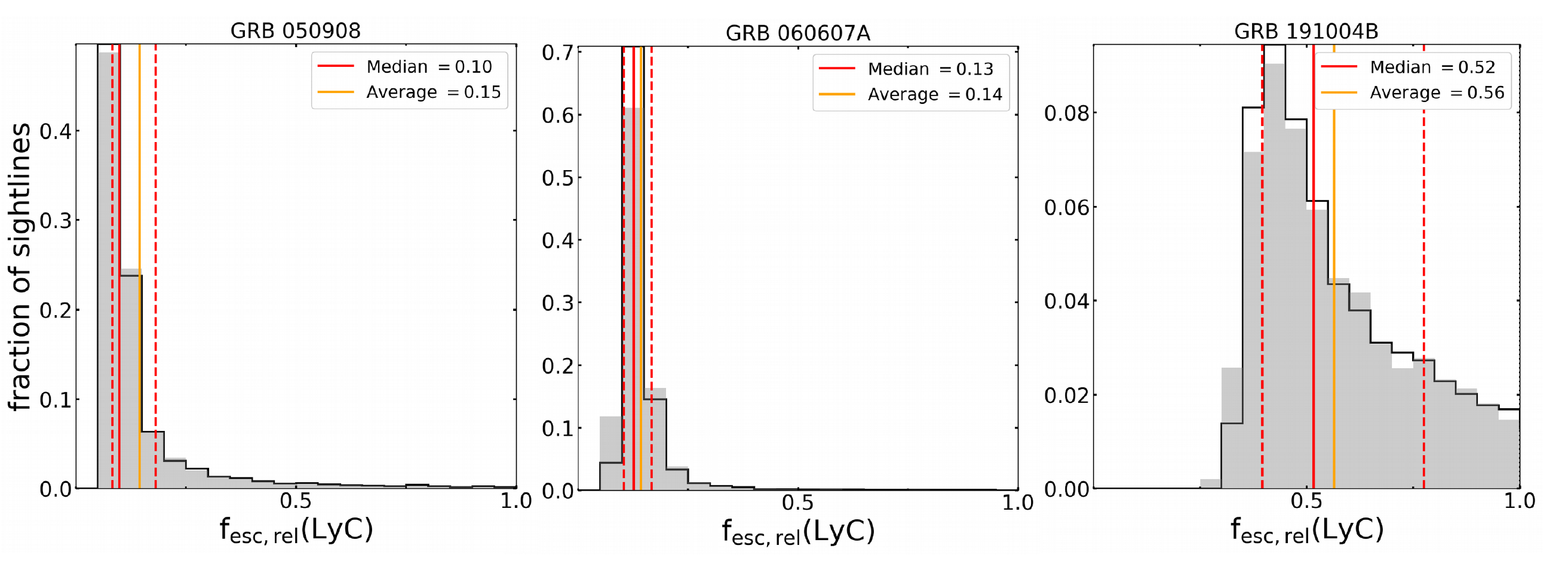}
 \caption{Distribution of relative escape fraction ($f_{\rm esc,rel}(\rm LyC)$) for GRB\,050908, GRB\,060607A and GRB\,191004B (from left to right) with and without uncertainty on the observed measure of $(f_{1500}/f_{900})^{\rm obs}$ from the afterglow spectrum (respectively gray and black distribution).
The high values with $f_{\rm esc,rel}(\rm LyC) > 1$ are non-physical and due to low IGM transmission, hence we restrict the statistics and axis range to 0-1 for $f_{\rm esc,rel}(\rm LyC)$.
For GRB\,191004B, 57\% of the gray distribution is at $f_{\rm esc,rel}(\rm LyC) \leq 1$ while the fraction is 92\% and 96\% for GRB\,050908 and GRB\,060607A, respectively. The solid red and orange lines in each panel represent the median and average value of this $f_{\rm esc,rel}(\rm LyC) \leq 1$ part of the distribution. The dashed red lines represent the corresponding 16th and 84th percentiles.
The $f_{\rm esc,~abs}$ are similar to the values quoted here for GRB\,050908 due to the null extinction hypothesis. For GRB\,191004B and GRB\,060607A the corresponding absolute values are derived in Sect.~\ref{IndirectMeasurement} and \ref{GRB_LyC_leakage}.
}
 \label{Distri_fesc}
\end{figure*}

The LyC escape fraction is the ratio of the observed flux below the Lyman limit, corrected for the
Milky Way extinction and the IGM transmission, and the intrinsic flux. Usually in galaxy studies this last quantity is inferred from the galaxy SED (and therefore it is model dependent), and the IGM transmission used is an average value obtained from line-of-sight simulations. 

Taking advantage of afterglow spectroscopy it is possible to determine $f_{\rm esc}$ \textit{(i):} directly using the \NHI~value derived by fitting the \Lya~absorption line or the envelope of the residual flux below the Lyman limit (see Sect.~\ref{DirectMeasurement}); 
or \textit{(ii):} through the $(f_{1500}/f_{900})^{\rm int}$ ratio and IGM simulation (see Sect.~\ref{IndirectMeasurement}), as commonly done in galaxy studies. 
The former case is a direct determination from the data.
The fact that the afterglow continuum is a simple power-law is an advantage (in both cases) because the intrinsic $f_{900}$ is much less model dependent than in galaxy studies.

\subsection{Direct determination} \label{DirectMeasurement}

Following \cite{Siana2007}, we define $f_{\rm esc}$ as the fraction of emitted Lyman continuum photons that escapes into the IGM. This corresponds to the formula,
\begin{equation}
f_{\rm esc} =  \exp[-\tau_{\rm LL}(\lambda)] ,
\label{fesc_tau}
\end{equation}

\noindent where $\tau_{\rm LL}$ is the optical depth at wavelengths below the Lyman limit and is given by (see also \citealt{Prochaska2010}),
\begin{equation}
\tau_{\rm LL}(\lambda)\approx \frac{ N_{\rm HI} }
{10^{17.2}\rm cm^{-2}}
\left(\frac{\lambda}{\lambda_{\rm LL}}\right)^3 ,
\label{tauLL}
\end{equation}

\noindent where $\lambda_{\rm LL}\approx912$~\AA~is the Lyman limit wavelength. 
In principle, we can directly use the value of \NHI~obtained by the fit of the \HI~absorption. In the case of GRB\,191004B, the \HI~column density belongs to the flat part of the \HI~curve of growth, therefore it cannot be determined precisely. Considering the value and errors determined in Sect.~\ref{absorptions}, we find $ f_{\rm esc} = 0.43^{+0.22}_{-0.24}$. The large error range reflects the \NHI~uncertainties.

We can determine more precise \NHI~and $ f_{\rm esc}$ values in the following way.
The observed afterglow spectrum blueward of the Lyman limit is the afterglow continuum (a power-law modulated by extinction, i.e., the solid line in Fig.~\ref{GRB191004B_powerLaw}), heavily absorbed by the neutral hydrogen in the host galaxy (Eq.~(\ref{tauLL})), and furthermore absorbed to a much lesser extent by Lyman alpha forest at the corresponding redshift. From the shape of the observed spectrum and the modeled afterglow continuum we can use Eq.~(\ref{tauLL}) to directly determine the 
opacity in the host galaxy's line-of-sight by varying the \NHI~(see Fig.~\ref{figFynbo}). Because the observed spectrum is partly absorbed by the Lyman alpha forest we don't aim to match the spectral continuum but instead its
envelope. The best match is obtained with \lognhicm~$=17.20\pm0.15$, which corresponds to
$\langle f_{\rm esc}\rangle = 0.43^{+0.12}_{-0.13}$. The errors reflect the uncertainty on the absorption due to the forest. 
If we take into account the effect of dust extinction on the afterglow continuum, we obtain $\langle f_{\rm esc,~abs}\rangle = 0.35^{+0.10}_{-0.11}$.

We can also determine a lower limit to $f_{\rm esc}$ and \NHI~from the observed continuum over the same wavelength range, as it corresponds to the flux of the escaping photons dimmed by the IGM opacity. We obtain \lognhicm~$<~17.4$, $f_{\rm esc}~>~0.25$, and $f_{\rm esc,~ abs}~>~0.22$ taking dust extinction into account.

\subsection{Constraints from IGM simulations} \label{IndirectMeasurement}

In galaxy studies, the common way to estimate the fraction of escaping ionizing photons is by calculating the ratio between the observed fraction of escaping LyC photons (900~\AA), corrected for IGM transmission, relative to the fraction of escaping non-ionizing photons (1500~\AA) \citep{Steidel2001, Siana2007}. The relative escape fraction can be expressed as:
\begin{equation}
 f_{\rm esc,rel} =\frac{ (f_{1500}/f_{900})^{\rm int} }{ (f_{1500}/f_{900})^{\rm obs}\ T_{900}^{\rm IGM}} ,
 \label{fescrel}
\end{equation} 

\noindent where $(f_{1500}/f_{900})^{\rm int}$ is the intrinsic flux density ratio, $(f_{1500}/f_{900})^{\rm obs}$ is the observed flux density ratio and $T_{900}^{\rm IGM}=e^{-\tau_{\rm IGM}}$ is the IGM transmission factor at 900~\AA~along the sightline, and $\tau_{\rm IGM}$ the line-of-sight opacity to the IGM for LyC photons. As shown in \cite{Siana2007}, this formula is equivalent to Eq.~(\ref{fesc_tau}).

The absolute escape fraction, $f_{\rm esc,abs}$ is the ratio of the escaping to intrinsic ionizing flux density. Knowing the dust attenuation $A_{1500}$, $f_{\rm esc,abs}$ can be written as \citep[e.g.,][]{Inoue2005, Siana2007},
\begin{equation}
 f_{\rm esc,abs} = f_{\rm esc,rel}\ 10^{-0.4\ A_{1500}} \, .
 \label{fescabs}
\end{equation} 

\noindent The $(f_{1500}/f_{900})^{\rm int}$ ratio is very difficult to constrain observationally
and is usually estimated by using spectral synthesis models \citep[e.g.,][]{Bruzual2003}. Therefore, this value has generally significant uncertainties due to the assumptions on which these models rely, stellar population age, metallicity, star-formation history and IMF. In our case, we have the possibility to directly measure the intrinsic flux density ratio as the intrinsic afterglow emission corresponds to a single power-law. 
This power-law is estimated as described in Sect.~\ref{extinction} and extrapolated to the ionizing UV domain. We then calculate the average intrinsic flux density in the same rest-frame range as for the observed flux densities ($f_{1500}^{\rm obs}$ and $f_{900}^{\rm obs}$, Sect.~\ref{LyC}). 
We finally derive an intrinsic flux ratio of $(f_{1500}/f_{900})^{\rm int} = 0.43$ from the best fit corrected for extinction (dashed orange curve in Figure \ref{GRB191004B_powerLaw}).

To recover the $T_{900}^{\rm IGM}$ factor, we simulated a large number of sightlines using observational constraints on 
the properties of the IGM probed in intergalactic absorbers. We used the same Monte Carlo (MC) simulation as in \citet{Japelj2017} and described in \citet{Vanzella2015} to generate 10\,000 line-of-sight transmissions from a source at $z=3.5$. 
For each simulated sightline we calculate the average transmission over the same rest-frame range as used for $f_{1500}^{\rm obs}$ and $f_{900}^{\rm obs}$, we then correct the observed ratio $(f_{1500}/f_{900})^{\rm obs}$ for this quantity and calculate $f_{\rm esc,rel}(\rm LyC)$.
In order to take into account the uncertainty on $(f_{1500}/f_{900})^{\rm obs}$, for each of the 10\,000 sightlines we randomly select a value in a normal distribution defined by its measurement (center) and its error (width). The resulting distribution of $f_{\rm esc,rel}$ is shown in grey in Fig.~\ref{Distri_fesc}.
In the case of GRB\,191004B, the best fit of the afterglow provides an estimate of the extinction and hence the possibility to calculate the absolute escape fraction. 
In the following, we report the absolute values derived using Eq.~(\ref{fescabs}) but for a comparison we also provide the relative ones, which do not rely on a dust-extinction model (see Sect.~\ref{extinction}).

For the estimation of the escape fraction we proceed similarly to what is done in \citet{Shapley2016} in their study of the direct detection of LyC emission from a galaxy at $z$\,$\sim$\,3.
We only keep the lines of sight resulting in an escape fraction of $f_{\rm esc} \leq 1$, since the values higher than one are not physical. 
We find that 57\% of the initial 10\,000 sightlines are at $f_{\rm esc} \leq 1$ 
and we measure that 95\% of the distribution is at $f_{\rm esc,~abs} \geq 0.31$ ($f_{\rm esc,~rel} \geq 0.35$). 
The median escape fraction is $f_{\rm esc,~abs} = 0.46_{-0.11}^{+0.23}$ ($f_{\rm esc,~rel} = 0.52_{-0.12}^{+0.26}$), with an average value of $\langle f_{\rm esc,~abs}\rangle = 0.50$ ($\langle f_{\rm esc,~rel}\rangle = 0.56$), see also Table \ref{Tab_fesc}.

\section{LyC leakage among LGRB host galaxies} \label{GRB_LyC_leakage}

\begin{table}[]
\begin{center}
\caption{Escape fraction of ionizing photons along the line of sight for the three GRBs.}
\label{Tab_fesc}
\centering
\resizebox{\columnwidth}{!}{%
\begin{tabular}{ c c c c c } \hline \hline
GRB & $z_{\rm syst}$    & $f_{\rm esc,abs~(rel)}^{\rm HI}$ &  $f_{\rm esc,abs~(rel)}$ & $f_{\rm esc,abs}^{\rm lim}$ \T\B \\
\hline
\rule[0.2cm]{0cm}{0.2cm}050908   & 3.3467       &  $0.08_{-0.04}^{+0.05}$ (*)    &  $0.10_{-0.02}^{+0.08}$ (*)                                   & $>0.07$ \B \\
060607A  & 3.0749       &  $0.20^{+0.05}_{-0.05}$ ($0.45^{+0.13}_{-0.13}$)   &  $0.08_{-0.02}^{+0.03}$  ($0.13_{-0.03}^{+0.04}$)             & $>0.05$ \B \\
191004B  & 3.5055       &  $0.35^{+0.10}_{-0.11}$ ($0.43^{+0.12}_{-0.13}$)   &   $0.46_{-0.11}^{+0.23}$ $\Big(0.52_{-0.12}^{+0.26}\Big)$     & $>0.31$ \B \\
\hline
\end{tabular}%
}
\tablefoot{The columns correspond to the name of the GRB; 
 $z_{\rm syst}$: the systemic redshift of the host galaxy; 
 $f_{\rm esc,abs~(rel)}^{HI}$: the absolute (abs) and relative (rel) escape fraction determined from the estimation of the \HI~column density, see Sect.~\ref{DirectMeasurement};
 $f_{\rm esc,abs~(rel)}$: the absolute (abs) and relative (rel) lower limit of the escape fraction of ionizing photons. The relative value is given between brackets when a non-zero extinction is considered, otherwise (*) means same value as the absolute one (see Sect.~\ref{IndirectMeasurement});
 $f_{\rm esc,abs}^{lim}$: the absolute lower limit of the escape fraction of ionizing photons (see Sect.~\ref{IndirectMeasurement}).}
\end{center}
\end{table}

\subsection{Other known LyC-leaking LGRB hosts}

To date, in addition to GRB\,191004B presented in this paper, there are only two more LGRBs for which direct LyC emission was detected, namely GRB\,050908 and GRB\,060607A. Their afterglow spectra have already been presented in \citet{Fynbo2009} and briefly discussed in \citet{Tanvir2019}. Here, we summarize the main known features of these LGRB afterglows and host galaxies and derive their $f_{\rm esc}(\rm LyC)$, following the procedures of Sect.~\ref{fesc}. We stress that, apart from their afterglow optical spectrum showing low-\HI~column densities associated with the LGRB host galaxy, the general properties of these LGRBs (prompt emission, energetic, afterglow light curves) are similar to that of the general LGRB population \citep[see, e.g.,][]{Kann2010}.

GRB\,050908 occurred at $z=3.3467$. Its afterglow spectrum was observed with VLT/FORS1 \citep{Fynbo2009}, KeckII/Deimos and Gemini North/GMOS \citep{Chen2007}, and showed very strong high- and low-ionization absorption lines at the GRB redshift. In this paper we use the spectrum obtained by VLT/FORS. 
In addition to the absorptions associated with the GRB host galaxy, two strong \ion{Mg}{II} intervening absorbers are identified in the GRB afterglow spectrum at $z=0.6915$ and 1.4288. The emission counterparts of these absorbers have been identified by \citet{Schulze2012} a few tens of kpc away from the GRB host. Deep {\it HST-F775W} images of the field allowed the detection of the GRB host galaxy with $m_{AB}=27.67$ mag \citep{Hjorth2012, Blanchard2016}. 
The VLT/FORS spectrum covers the ionizing domain at wavelengths $>818$~\AA~rest-frame. A sufficient signal-to-noise ratio is only reached down to 880~\AA. We measure the ionizing ($f_{900}^{\rm obs}$) and non-ionizing UV ($f_{1500}^{\rm obs}$) flux density from this spectrum and find a $(f_{1500} / f_{900})^{\rm obs}$ ratio of $6.9 \pm 0.5$. 
Considering a null extinction ($A_{\rm V}=0$ mag), according to the results published in \citet{Vergani2008}, we fit the afterglow spectrum of GRB\,050908 with a simple power-law and
following the method described in Sect.~\ref{DirectMeasurement}, we derive $(f_{1500}/f_{900})^{\rm int} = 0.35$ with $f_{900}^{\rm int}$ measured over 880-910~\AA. Similarly to GRB\,191004B we can derive the absolute escape fraction of ionizing photons which is equal to the relative one in this case, since no extinction correction is applied.
The distribution of $f_{\rm esc,rel}(\rm LyC)$ is presented in Fig.~\ref{Distri_fesc} and shows that 95\% of the distribution is at $f_{\rm esc,~abs} \geq 0.07$ and 92\% with $f_{\rm esc,~abs} \leq 1$. 
The median escape fraction for the latter 92\% is $0.10_{-0.02}^{+0.08}$, with an average value of $\langle f_{\rm esc,~abs}\rangle = 0.15$. 
This estimation is fully consistent with the value of $f_{\rm esc}=0.08^{+0.05}_{-0.04}$ derived by \cite{Tanvir2019} from the \HI~column density (\lognhicm$=17.6$), following the method presented in Sect.~\ref{DirectMeasurement}.

GRB\,060607A occurred in a galaxy at $z=3.0749$. It has the lowest \HI~column density observed among the LGRB afterglow sample (\lognhicm $~=16.95\pm0.03$; \citealt{Tanvir2019}) and, like GRB\,191004B, the only metal absorption lines at the GRB redshift are those of \ion{C}{iv} and \ion{Si}{iv}.
The VLT/UVES spectrum of its afterglow shows a very complex line of sight, rich in intervening absorbers \citep{Fox2008, Prochaska2008, Fynbo2009}. There is evidence of LyC emission over $\sim$16~\AA~below the LL, likely absorbed blueward from an intervening system. It is not firmly possible to exclude that the apparent LyC emission is due to an interloper. Nonetheless, 
deep {\it HST-F775W} imaging observations are available for the GRB\,060607A field (see \citealt{Blanchard2016}) and no galaxy is detected at $m_{AB}>28.9$ mag over a radius of $\sim$15~kpc from the afterglow position.
From the UVES spectrum, we measure the ionizing ($f_{900}^{\rm obs}$) flux density over the 16~\AA~where LyC emission is detected and we find a $(f_{1500} / f_{900})^{\rm obs}$ ratio of $5.0\pm0.6$.
The determination of dust extinction on this line of sight is not trivial as the afterglow experienced some rebrightening during the X-ray observations and UVES spectroscopy. Following the same prescriptions as in Sect.~3.3, we find a quite high extinction with $A_{\rm V} = 0.13\pm0.04$ mag for the Small Magellanic Cloud extinction curve \citep[see also][]{Kann2010}.
Following the same method as for the previous cases, we derive 
$(f_{1500}/f_{900})^{\rm int} = 0.42$.
The distribution of $f_{\rm esc,rel}(\rm LyC)$ is presented in Fig.~\ref{Distri_fesc} and shows that 95\% of the distribution is at $f_{\rm esc,~abs} \geq 0.05$ ($f_{\rm esc,~rel} \geq 0.09$) and 96\% with $f_{\rm esc,rel} \leq 1$. 
The median escape fraction for the latter 96\% is $f_{\rm esc,~abs} = 0.08_{-0.02}^{+0.03}$ ($f_{\rm esc,~rel} = 0.13_{-0.03}^{+0.04}$) with an average value of $\langle f_{\rm esc,~abs}\rangle = 0.08$ ($\langle f_{\rm esc,~rel}\rangle = 0.14$).
The value derived from the \HI~column density, following the method presented in Sect.~\ref{DirectMeasurement}, is $f_{\rm esc}= 0.45 \pm 0.13$ and $f_{\rm esc,~abs}=0.20\pm0.05$, with \lognhicm $~=17.1\pm0.15$ (in agreement with the column density reported in the literature). The difference of the $f_{\rm esc}$ values determined by the two methods could be due to a higher $T_{900}^{\rm IGM}$ average from simulations, as this line of sight is particularly rich in absorbing systems.
We stress that both $f_{\rm esc,~abs}$ values are tentative due to the difficulties in determining the dust extinction.
The $f_{\rm esc}$ results for the three GRBs are summarized in Table \ref{Tab_fesc}.

\subsection{Statistical estimation of the ionizing escape fraction in LGRB host galaxies} \label{Stat_fesc}

The \HI~column density of \lognhicm $~= 17.2 \pm 0.15$ derived for GRB\,191004B corresponds to one of the lowest \NHI~measured for an LGRB host galaxy.
Adding this case to the sample studied in \citet{Tanvir2019} we re-evaluate the average escape fraction derived from the compilation of the 141 column densities available to-date. To do so, we use the formula described in Sect.~\ref{DirectMeasurement} (Eq.~(\ref{fesc_tau})) averaged over the full sample and we find $\langle f_{\rm esc}\rangle = 0.007 $ instead of 0.005 as presented in \citet{Tanvir2019}. Again following the method used in \citet{Chen2007b} and \citet{Tanvir2019} we performed a bootstrap resampling by allowing replacement of the 141 sightlines in the above sample. From the $10^6$ \NHI~distributions simulated with this method and taking into account scatter produced by the uncertainty on each data point, we find a 98\% c.l upper limit of $\langle f_{\rm esc} \rangle < 0.020$. This new estimation increases somewhat the result found in \citet{Tanvir2019} but is still consistent with their $\langle f_{\rm esc}\rangle$ upper limit in the range between 0.015 - 0.02.

\section{Modeling of the \Lya~emission line} \label{Lya_fitting}

\begin{figure}
\centering
\includegraphics[width=\hsize]{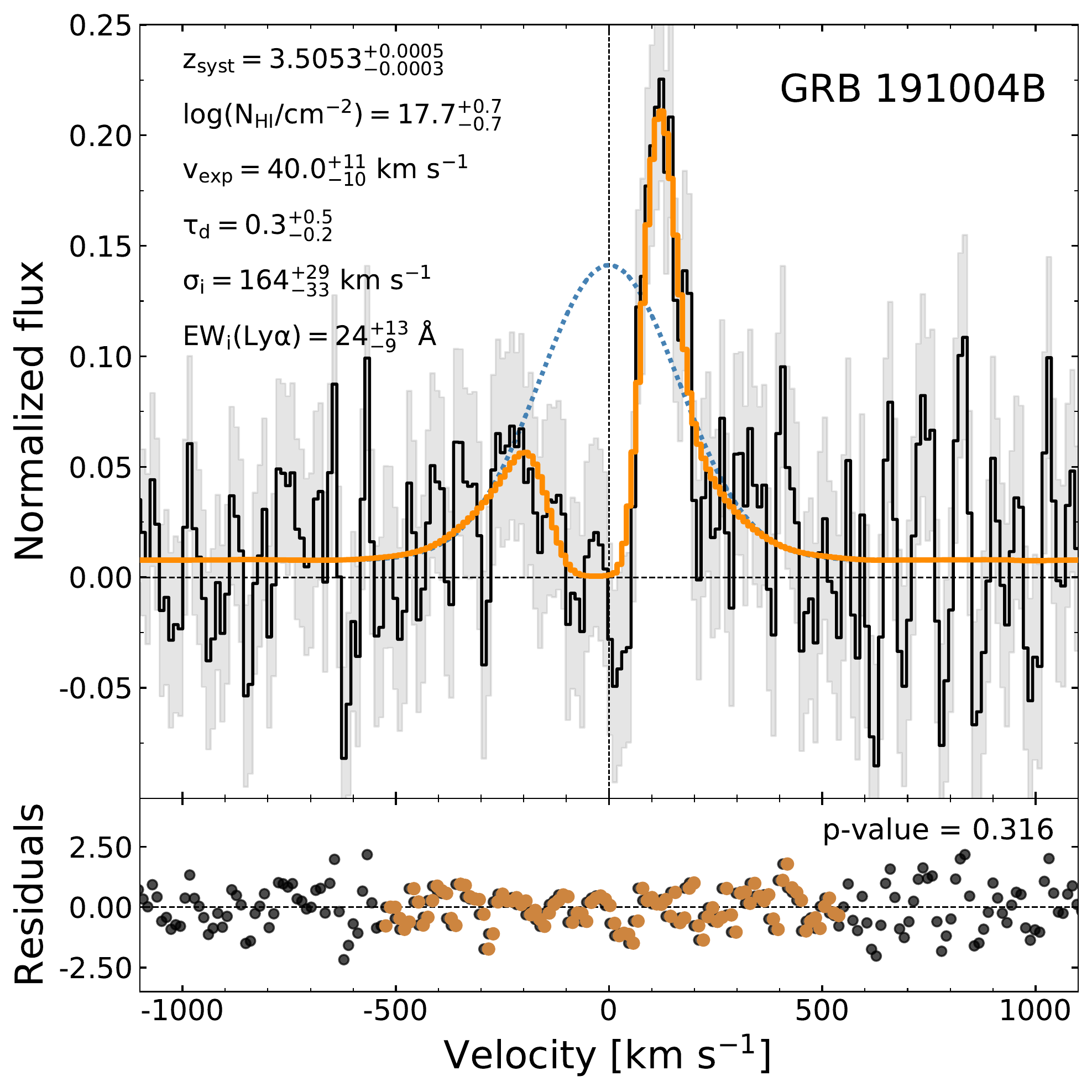}
 \caption{Best fit of the GRB\,191004B \Lya~line, observed in the X-shooter host-galaxy spectrum, with unconstrained shell models. 
 In the top panel, the solid black line corresponds to the data with their error bars (grey), in dashed orange the best fit and in dotted blue the intrinsic \Lya~emission predicted by the model. In the top left corner, we show the best-fit parameters.
 In the bottom panel, the dots correspond to the normalized residuals between the observation and the model. The orange ones are for the \Lya~profile only with the corresponding p-value for the Shapiro-Wilk test.
 }
 \label{fit_Lya}
\end{figure}

The non detection of nebular emission lines in the spectra of GRB\,191004B prevents firm determination of the exact redshift of the host galaxy. 
We can, however, use the empirical correlation found by \citet{Verhamme2018} between the FWHM and the velocity shift of the peak of the \Lya~line.
We derive a predicted redshift of $z_{\rm predict} = 3.5060 \pm 0.0008$. 
This value is fully consistent with the \HI~component seen in absorption at $z=3.5055$.

Due to its resonant nature, the \Lya~radiation produced by active star-forming regions carries the signature of the physical properties of the neutral gas where the photons scatter before escaping.
In that sense, the detection of the \Lya~line gives us a unique chance to investigate if the properties probed by the GRB afterglow lines of sight are similar to the ones of the material probed by the \Lya~emission.
To do this comparison, we perform a shell-model fitting of the observed \Lya~line using an improved model grid and process originally described in \citet{Gronke2015}. 
The shell model is a commonly used model (first introduced by \citealt[][]{Ahn2003}, also see \citealt{Verhamme2006}) that consists of a central source emitting a continuum and a Gaussian \Lya~line defined by an intrinsic width and equivalent width ($\sigma_{\rm i}$, $EW_{\rm i}$). This source is surrounded by a shell of neutral hydrogen and dust described by four parameters: a radial expansion velocity $v_{\rm exp}$; a \HI~column density \NHI; an effective temperature of the gas {\it T}; an optical depth of dust close to \Lya~wavelength $\tau_{d}$. 
We fit the X-shooter host-galaxy data leaving these parameters free in the grid range values described in \citet{Gronke2015}.
As the final fitting result is sensitive to the redshift of the emitting source, we also leave the systemic redshift as a free parameter with a Gaussian prior of $z_{\rm syst}=3.5060 \pm 0.01$, based on the estimation derived above. 
Note, however, that the prior on $z_{\rm syst}$ is very wide, and thus, essentially a free parameter. 

We compare the observed \Lya~profile to the best-fit model in Fig.~\ref{fit_Lya}.
We find that the best-fit profile describes the observation very well.
Indeed, the normalized residuals between the observed and modeled \Lya~line (brown dots in Fig.~\ref{fit_Lya}) provides a Shapiro-Wilk coefficient of 0.985 and a p-value $= 0.316$, therefore consistent with Gaussian noise.
The best-fit results show a small blue component at -210~\kms, consistent with the spectrum. 
This blue component is only detected at a 2.5$\sigma$ confidence level, with a flux of $F(\rm Ly\alpha, blue) = (0.5 \pm 0.2) \times 10^{-17}$~\ergscm. If the blue bump is real, the separation between the blue and red peaks is 323~\kms.

The best-fit parameters are consistent with the observed values,
\textit{(i):} the systemic redshift $z_{\rm syst}=3.5053^{+0.0005}_{-0.0003}$ is in good agreement with the value of the reddest \HI~absorption component seen at $z=3.5055$ and also with $z_{\rm predict} = 3.5060 \pm 0.0008$;
\textit{(ii):} the low expansion velocity of the gas $v_{\rm exp}=40.0^{+12.0}_{-11.0}$~\kms~is consistent with the velocity of the reddest \HI~component which also is the richest in neutral hydrogen (cf. Fig.~\ref{fig:fit_abs_GRB191004B});
\textit{(iii):} the predicted \HI~column density, \lognhicm $~= 17.7^{+0.7}_{-0.7}$ and  
\textit{(iv):} the dust optical depth $\tau_{d}=0.3^{+0.5}_{-0.2}$ corresponding to $A_{\rm V} = 0.04^{+0.07}_{-0.04}$ mag are fully consistent with the observational values derived from the afterglow spectrum (see Sect.~\ref{extinction} and \ref{absorptions});
\textit{(v):} the intrinsic properties of \Lya~and Balmer lines are expected to be similar if we assume both transitions to be formed by recombination in the same regions of the galaxy.
Here, we cannot constrain the intrinsic properties of the \Lya~source ($\sigma_i$, $EW_i$) as we do not observe the Balmer lines. 
Nevertheless, by considering the recombination factor 8.7 (Sect.~\ref{LyaE}) and the \Ha/\Hb~flux ratio of 2.86 (Sect.~\ref{Emissions}),  
the intrinsic \Lya~properties predicted by the best-fit model ($\sigma_i=164^{+29}_{-33}$~\kms, $EW_i(\rm Ly\alpha)=24^{+13}_{-9}$~\AA) imply a rest-frame $EW(\rm H\beta)=1.0^{+0.6}_{-0.4}$~\AA~which is fully consistent with its non detection in the X-shooter spectrum ($EW(\rm H\beta) <3.0$~\AA, see Sect.~\ref{Emissions}).

We stress that, while it is well known that the shell-model can reproduce observed \Lya~spectra 
extremely well \citep[e.g.,][]{Gronke2017}, its oversimplification of realistic \HI~configurations in and around galaxies is a source of debate in the literature parameters 
\citep[e.g.,][]{Gronke2017_607A, Orlitova2018}.

\section{Discussion} \label{discussion}

\subsection{Comparison with \Lya-emitter LGRBs (LAE-LGRBs)} \label{CompLAE}

The \Lya~line profile of GRB\,191004B is typical of what is observed for known LAE-LGRBs.
It shows a main peak redshifted by $113 \pm 30$ \kms~which is just below the range of values (150-750 \kms) determined for the known LAE-LGRB host-galaxies in \citet{MilvangJensen2012}.
The low-velocity shift of the line associated with LyC leakage is consistent with the theoretical predictions of \citet{Verhamme2015}. Indeed, they found that the classical asymmetric redshifted profile of the \Lya~line should be shifted by less than 150 \kms~in the case of LyC leakage.
This was also observationally confirmed in \citet{Verhamme2017} for low-redshift LyC emitters.
The host galaxy has a \Lya~luminosity ($L_{\rm Ly\alpha} = (1.18~\pm~0.18) \times 10^{42}$ erg~s$^{-1}$) and a UV magnitude lower limit ($M_{\rm UV}>-19.2$ mag) found to be close to the median values of the LAE-LGRB host-galaxies sample, respectively, $L_{\rm Ly\alpha} = 1.7^{+3.5}_{-0.9} \times 10^{42}$ erg~s$^{-1}$ \citep{Fynbo2003, Jakobsson2005, MilvangJensen2012} and $M_{\rm UV} = -19.9^{+1.4}_{-1.0}$ mag (\citealt {Tanvir2019}).
However, GRB\,191004B is the only LAE-LGRB for which LyC leakage has been detected (over the eight LAE-LGRBs with a spectrum covering the spectral range beyond the Lyman limit) or inferred from the \HI~column density probed by the afterglow. It is in fact the only LAE-LGRB having \lognhicm $~<18$, over the twenty-two LAE-LGRBs with an afterglow spectrum covering the \Lya~absorption spectral range.
Indeed, LGRBs usually probe dense \HI~column densities.
Among the 140 LGRB lines of sight of the sample of \citet{Tanvir2019}
only two have \lognhicm $~<18$, namely GRB\,050908 (\lognhicm$=$17.6) and  GRB\,060607A (\lognhicm$=$16.95).
Consistently, together with GRB\,191004B, they also are the only cases showing non-zero afterglow continuum emission below the Lyman limit.
Intriguingly they do not show any trace of \Lya~emission above a 3$\sigma$ flux upper limit of $F(\rm Ly\alpha) = 6.4 \times 10^{-18}$ \ergscm~and $F(\rm Ly\alpha) = 7.3 \times 10^{-18}$ \ergscm, respectively \citep{MilvangJensen2012}. We should consider, however, that those are extremely faint galaxies.
Even if no firm conclusion can be drawn from such a small sample, the fact that 
the highest escape fraction is obtained for GRB\,191004B, which is also the only LAE, 
seems to support the idea that strong LyC emission is correlated with \Lya~emission.
Inversely, substantial \Lya~emission is not necessarily associated with LyC leakage, as GRB\,191004B is the only LyC leaker among eight LAE-LGRBs of comparison. 

LyC and \Lya~escape are expected to be somehow correlated \citep[see, e.g.,][]{Verhamme2015,Verhamme2017,Dijkstra2016,Kakiichi2019} since the escape of both radiations is facilitated by low-\NHI~sightlines. 
Nevertheless, there exist physical phenomena depending mainly on the density of the \HI~gas and the presence of low density paths in the medium, that can boost the \Lya~against LyC escape, or inversely \citep[see, e.g.,][Sect.~3.2]{Ji2020}.
Indeed, recent studies such as the one of \citet{Bian2020} show that strong \Lya~emission is not necessarily a good indicator of LyC leakage \citep[see also e.g.,][]{Guaita2016, Grazian2017}. Observationally, the \Lya~EW does seem to correlate with LyC escape as shown by \citet{Steidel2018}. There are no deep enough observations to put significant \Lya~EW limits for GRB\,050908 and GRB\,060607A hosts, but in general,
galaxies with comparably high LyC $f_{\rm esc}$ as GRB\,191004B have higher \Lya~EW$_0$ (e.g.  \citealt{Izotov2016b,Izotov2018a,Vanzella2016}). A relation between the separation of the observed \Lya~emission blue and red peaks and the LyC escape fraction has also been found by e.g. \cite{Verhamme2015, Verhamme2017} and \citet{Izotov2018b} for low-redshift LyC leaking LAEs. The separation of $\sim$300~\kms~found for GRB\,191004B would correspond to $f_{\rm esc}$\,$\sim$10\%, thus, lower than the value we determined. 
In our case, we are probing the LyC leakage along the GRB line of sight. 
Therefore, it is possible that the average LyC leakage of the galaxy and the \NHI~through which \Lya~photons escape are different than those probed by the GRB afterglow. A higher average \NHI~value, as the best value of the shell-model fitting (\lognhicm~$=17.7$), would imply an $f_{\rm esc}$(LyC) of 6.9\%, in better agreement with the \Lya~EW$_0$ and peak separation relations.
In this case, our results could support a medium with low density holes and the escape would not be homogeneous. This is a more realistic scenario, as the ISM in galaxies is not homogeneous.
We would like to point out also that the relations above have not been tested for faint, high-redshift galaxies as those studied in this paper. \cite{Steidel2018} explore  brighter galaxies, whereas \cite{Fletcher2019} study systems with \Lya~EW$_0>15$~\AA.

\subsection{Comparison with known LyC emitters}

\begin{figure}
\centering
\includegraphics[width=\hsize]{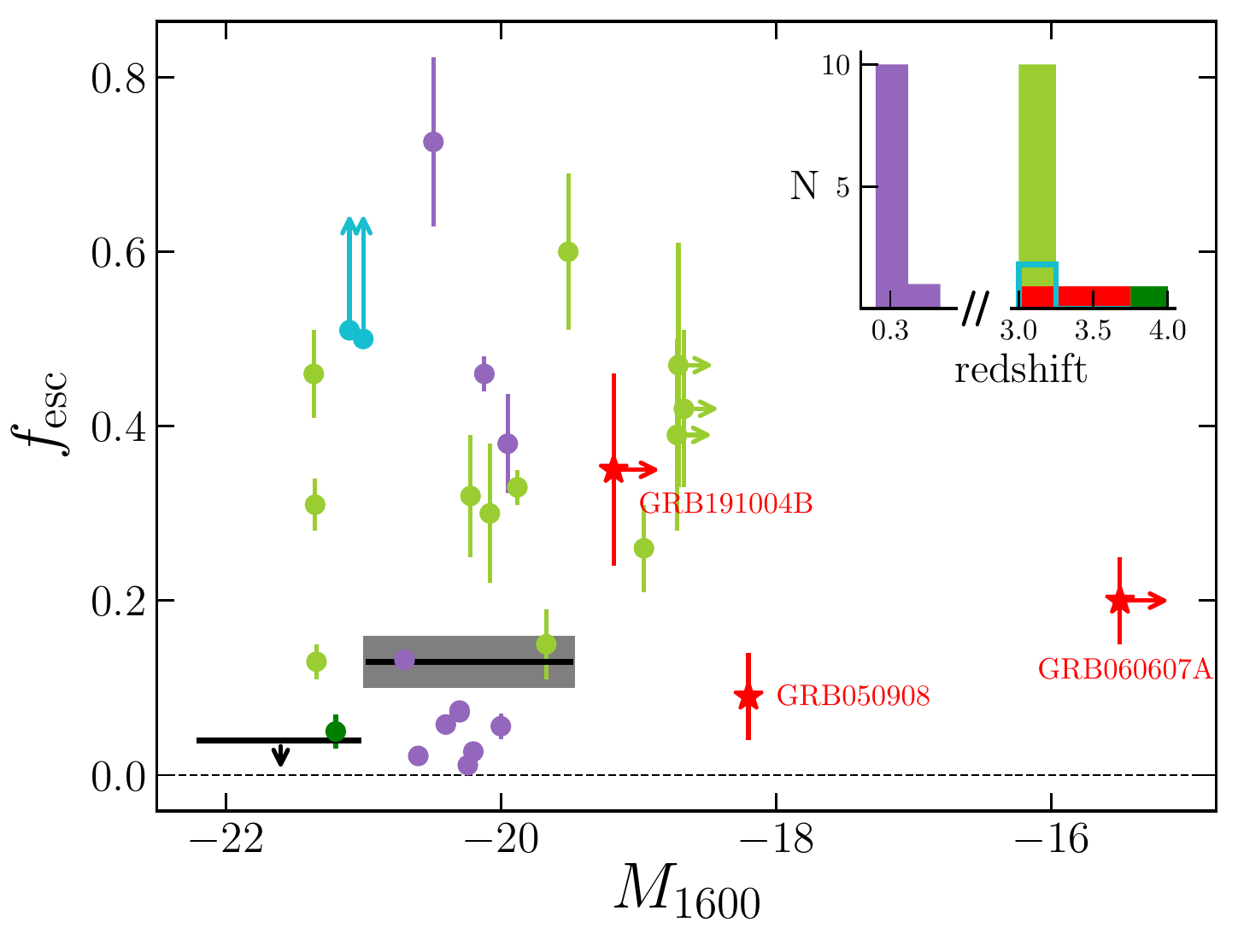}
 \caption{Collection of measurements of absolute escape fractions of known LyC emitters as a function of galaxy UV luminosity. The LGRB lines of sight studied in this paper are shown with red stars. Other data points are taken from \citet{Fletcher2019} ($z$\,$\sim$\,3, light green), \citet{Izotov2016b,Izotov2018a,Izotov2018b} ($z$\,$\sim$\,0.3, violet), \citet{Ji2020} ($z$\,$\sim$\,3.8, dark green) and \citet{Vanzella2016} and \citet{Shapley2016} ($z$\,$\sim$\,3.2, blue). Black lines indicate the limit and the average as provided by the $z$\,$\sim$\,3 sample study of \citet{Steidel2018}. The inset shows the redshift distribution of the systems.
 }
 \label{fesc_M1600}
\end{figure}

We collected absolute escape fraction measurements from the literature, in particular from galaxies lying at $z \gtrsim 3$. The comparison between our LGRB measurements and those from the literature is presented in Fig.~\ref{fesc_M1600} as a function of galaxy UV luminosity. 

The data in the plot is not homogeneous, owing to the difficulty of finding LyC emitters and measuring their escape fractions. The IGM transmission $T_{900}^{\rm IGM}$ depends on the assumed model; \citet{Steidel2018} and \citet{Shapley2016} use a combined transmission of the IGM and circumgalactic medium (CGM), resulting in on average lower transmission than our model (and that of \citealt{Vanzella2016}). \citet{Fletcher2019} adopt an average transmission at $z$\,$\sim$\,3. If they used a similar probabilistic approach as in this work, the measurements from their study would have lower values, and they would be only a lower limit. The choice of $(f_{1500}/f_{900})^{\rm int}$ also varies based on different galaxy population synthesis studies. Finally, we overplot the sample of $z$\,$\sim$\,0.3 green pea LyC emitters. Their low-redshift origin makes the measurements less affected by the uncertainties in $T_{900}^{\rm IGM}$ and $(f_{1500}/f_{900})^{\rm int}$.

Figure~\ref{fesc_M1600} shows that LGRBs can be powerful tools to probe LyC leakage from extremely faint high-redshift galaxies, making it possible to explore LyC leakage from galaxies beyond the current absolute magnitude limits of galaxy studies. 
The number of objects at $M_{\rm 1600}>-19.5$ mag is similar to those of galaxy studies and the power to access even fainter galaxies is evident.
LyC leakage investigations through LGRB afterglows do not suffer from galaxy apparent magnitude selection.
Furthermore, the results are obtained as a by-product of routinely performed observations for the GRB redshift determination, requiring a very small observing time (approximately one-three hours). 

We stress that there is a fundamental difference between our study, using LGRB afterglows, and the more common studies using galaxy observations. In our case, we look into the transparency of the galactic material to the ionizing radiation in one line of sight (toward the studied LGRB), while the other studies measure the averaged transparency of a whole galaxy facing us. Therefore, through LGRB afterglow we can probe directly the lines of sight through which LyC photons escape. 

The use of LGRBs to constrain the leakage of LyC photons can be a powerful way to understand what sources reionized the Universe.
Indeed, LGRBs can probe $f_{\rm esc}$ not only at higher redshift -- thus, requiring less extrapolation in terms of galactic properties -- but also (UV) fainter galaxies. 
This is particularly noteworthy since those galaxies are thought to be the main driver of reionization because 
\textit{(i)} models and analytical considerations show that their feedback is likely overruling gravity, thus, strong enough to clear channels through which LyC photons can escape \citep{Paardekooper2015,Cen2020}, and 
\textit{(ii)} the (increasing) steepness of the \Lya~luminosity function hints toward a very steep LyC luminosity function \citep{Gronke2015a,Dressler2015,Dijkstra2016}. 
A larger sample of LGRB afterglow spectra will allows us to test these theories.

\section{Conclusions}

We investigated LyC leakage in LGRB afterglow spectra through the observations 
of GRB\,191004B at $z=3.5055$ (presented here for the first time), GRB\,050908 ($z=3.0749$), and GRB\,060607A ($z=3.3467$). 
We found absolute escape fractions along their sightlines of $0.35^{+0.10}_{-0.11}$, $0.08^{+0.05}_{-0.04}$,$0.20^{+0.05}_{-0.05}$, respectively. By using simulations of the IGM opacity, we found similar results except for GRB\,060607A likely because of its very rich line of sight.

Thanks to the fact that the intrinsic afterglow emission corresponds to a single power-law, the determination of the intrinsic flux density ratio between the fraction of escaping ionizing and non-ionizing photons does not suffer from the uncertainties due to the assumptions behind galaxy spectral-synthesis models. Furthermore afterglow-based studies do not suffer from galaxy magnitude selection effects as galaxy studies.

The results presented here are by-products of routinely performed observations to determine the redshift of GRBs obtained using only approximately one-three hours each, compared to the many hours needed for galaxy-based studies to pre-select and observe such faint galaxies. They show that LGRBs can be powerful tools to study LyC leakage from faint, star-forming galaxies at high redshift. Indeed, the host galaxies of the LGRBs presented here all have $M_{\rm 1600}>-19.5$ mag, with the extreme case of the host of GRB\,060607A at $M_{\rm 1600}>-16$ mag. Such faint galaxies are very common at very high redshift. Their role in reionization is still debated, but their global ionizing photon budget may contribute significantly to the reionization process.

LGRBs explode in young star-forming regions, major contributors of the LyC photons inside galaxies. LGRB afterglows probe lines of sight originating from these regions, therefore, they shed light on the paths through which LyC photons escape.
Uniquely LGRBs also have the potential to combine the study of these lines of sight (through afterglow spectroscopy) with those of the global properties of their hosting galaxies, through photometric and spectroscopic observations once the afterglow has disappeared. The galaxies presented in this paper are faint and at the limit of current instrumentation, but they will be perfect targets for E-ELT and JWST. Future observations will allow us to determine their characteristics and to define indirect indicators to be used to find similar galaxies at the reionization epoch.

The current limitation of the LGRB contribution to reionization studies is the limited number of LGRBs with afterglow detections, especially at high redshift.
Future satellites such as {\it Gamow Explorer} \citep{White2020} and {\it THESEUS}\footnote{\url{https://www.isdc.unige.ch/theseus/}} (pre-selected as M5 European Space Agency mission), will largely improve this situation. Indeed the latter will allow the detection and redshift estimate of about 150 GRBs at $3<z<4$, during its three-year nominal duration \citep{Amati2018,Ghirlanda2015}. Furthermore, its detection of $\sim$100 LGRBs at $z>6$, combined with space and ground based observations, will allow us to directly study the galaxies that contribute to reionization.

\begin{acknowledgements}
This work is part of the {\it BEaPro} project (P.I.: S.D. Vergani) funded by the French {\it Agence Nationale de la Recherche} (ANR-16-CE31-0003). We thank Giancarlo Ghirlanda for providing useful information. JBV and SDV thank Anne Verhamme for useful discussions. 
SDV acknowledges financial support from the French Space Agency (CNES).
MG was supported by NASA through the NASA Hubble Fellowship grant HST-HF2-51409 and acknowledges support from HST grants HST-GO-15643.017-A, HST-AR-15039.003-A, and XSEDE grant TG-AST180036.
The Cosmic DAWN center is funded by the DNRF. JPUF thanks the Carlsberg foundation for support.
DBM acknowledges support from VILLUM FONDEN research grant 19054.
NRT acknowledges support from STFC via grant ST/N000757/1.
DAK acknowledges support from Spanish National Project research project
RTI2018-098104-J-I00 (GRBPhot).
The Pan-STARRS1 Surveys (PS1) and the PS1 public science archive have been made possible through contributions by the Institute for Astronomy, the University of Hawaii, the Pan-STARRS Project Office, the Max-Planck Society and its participating institutes, the Max Planck Institute for Astronomy, Heidelberg and the Max Planck Institute for Extraterrestrial Physics, Garching, The Johns Hopkins University, Durham University, the University of Edinburgh, the Queen's University Belfast, the Harvard-Smithsonian Center for Astrophysics, the Las Cumbres Observatory Global Telescope Network Incorporated, the National Central University of Taiwan, the Space Telescope Science Institute, the National Aeronautics and Space Administration under Grant No. NNX08AR22G issued through the Planetary Science Division of the NASA Science Mission Directorate, the National Science Foundation Grant No. AST-1238877, the University of Maryland, Eotvos Lorand University (ELTE), the Los Alamos National Laboratory, and the Gordon and Betty Moore Foundation.
This work has made use of data from the European Space Agency (ESA) mission
{\it Gaia} (\url{https://www.cosmos.esa.int/gaia}), processed by the {\it Gaia}
Data Processing and Analysis Consortium (DPAC,
\url{https://www.cosmos.esa.int/web/gaia/dpac/consortium}). Funding for the DPAC
has been provided by national institutions, in particular the institutions
participating in the {\it Gaia} Multilateral Agreement.
The NumPy \citep{vanderWalt2011}, SciPy \citep{2020SciPy} and matplotlib \citep{Hunter2007} packages have been extensively used for the preparation and presentation of this work.

\end{acknowledgements}

\bibliographystyle{aa}
\bibliography{refs}

\end{document}